\documentclass[11pt]{article}
\usepackage{amsmath}
\usepackage{amsfonts}
\usepackage{amssymb}
\usepackage{epsfig}
\usepackage{graphicx}
\usepackage{times}
\def\figskip{\vskip .5cm plus 3mm minus 2mm}
\def\hbar{\not{\hbox{\kern-2.3pt $h$}}}      
\begin{document}

\begin{titlepage}
%
September 2004 (revised January 2005)\hfill PAR-LPTHE 04/21\newline\null
\hfill ITEP-PH-3/2004\newline\null
\hfill hep-ph/0407268
\vskip 4cm
{\baselineskip 17pt
\centerline{
{\bf BINARY SYSTEMS OF NEUTRAL MESONS IN QUANTUM FIELD THEORY}}
}
\vskip .5cm
\centerline{B. Machet
        \footnote[1]{Laboratoire de Physique Th\'eorique et Hautes
Energies, UMR 7589, Unit\'e associ\'ee  au CNRS et aux
universit\'es Pierre et Marie Curie (Paris 6) et Denis Diderot (Paris 7);
postal address:  LPTHE tour 24-25, 5\raise 3pt \hbox{\tiny \`eme} \'etage,
          Universit\'e P. et M. Curie, BP 126, 4 place Jussieu,
          F-75252 Paris Cedex 05 (France)}
        \footnote[2]{machet@lpthe.jussieu.fr}
, V.A. Novikov
        \footnote[3]{ITEP, lab.~180, B. Cheremushkinskaya Ul. 25, 117218 Moscow
(Russia)}
        \footnote[4]{novikov@heron.itep.ru}
 and M.I. Vysotsky
        \footnotemark[3]
        \footnote[5]{vysotsky@heron.itep.ru} }

\vskip 1cm
{\bf Abstract:} Quasi-degenerate binary systems of neutral mesons of the kaon
type are investigated in Quantum Field Theory (QFT). General constraints cast by
analyticity and  discrete symmetries $P$, $C$, $CP$, $TCP$  on the propagator
(and on its spectral function) are deduced.
Its poles are the physical masses;
this unambiguously defines the propagating eigenstates.
It is diagonalized and its spectrum thoroughly investigated.
The role of ``spurious'' states, of zero norm at the poles, is emphasized,
in particular for unitarity and for the realization of  $TCP$ symmetry.
The $K_L-K_S$ mass splitting triggers a tiny difference between their  $CP$
violating parameters $\epsilon_L$ and $\epsilon_S$, without any
violation of $TCP$.
 A constant mass matrix like used in Quantum Mechanics (QM)
can only be introduced in a linear approximation to the inverse propagator,
which respects its analyticity and positivity properties;
it is however unable to faithfully describe all features of neutral
mesons as we determine them in QFT, nor to provide any sensible
parameterization of eventual effects of $TCP$ violation.
The suitable way to diagonalize the propagator makes use of a
bi-orthogonal basis; it is inequivalent to a bi-unitary transformation
(unless the propagator is normal, which cannot occur here).
Problems linked with the existence of different
``in'' and ``out'' eigenstates  are smoothed out.
We study phenomenological consequences of the differences between the QFT
and QM treatments; the non-vanishing of the semi-leptonic asymmetry
$\delta_S - \delta_L$, does not signal, unlike usually claimed, $TCP$
violation, while $A_{TCP}$ keeps vanishing when $TCP$ is realized.
We provide expressions invariant by the rephasing of $K^0$ and $\overline{K^0}$.

\smallskip

\vfill
\vbox{
\begin{center}
\hskip 1cm\includegraphics[height=2truecm,width=2truecm]{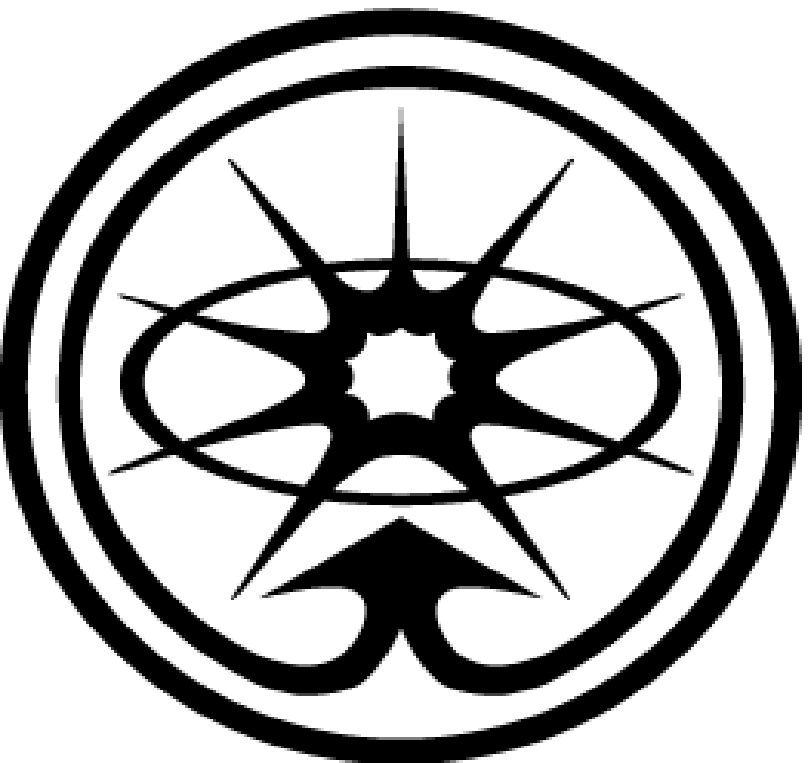}
\hskip 3cm
\includegraphics[height=2truecm,width=5truecm]{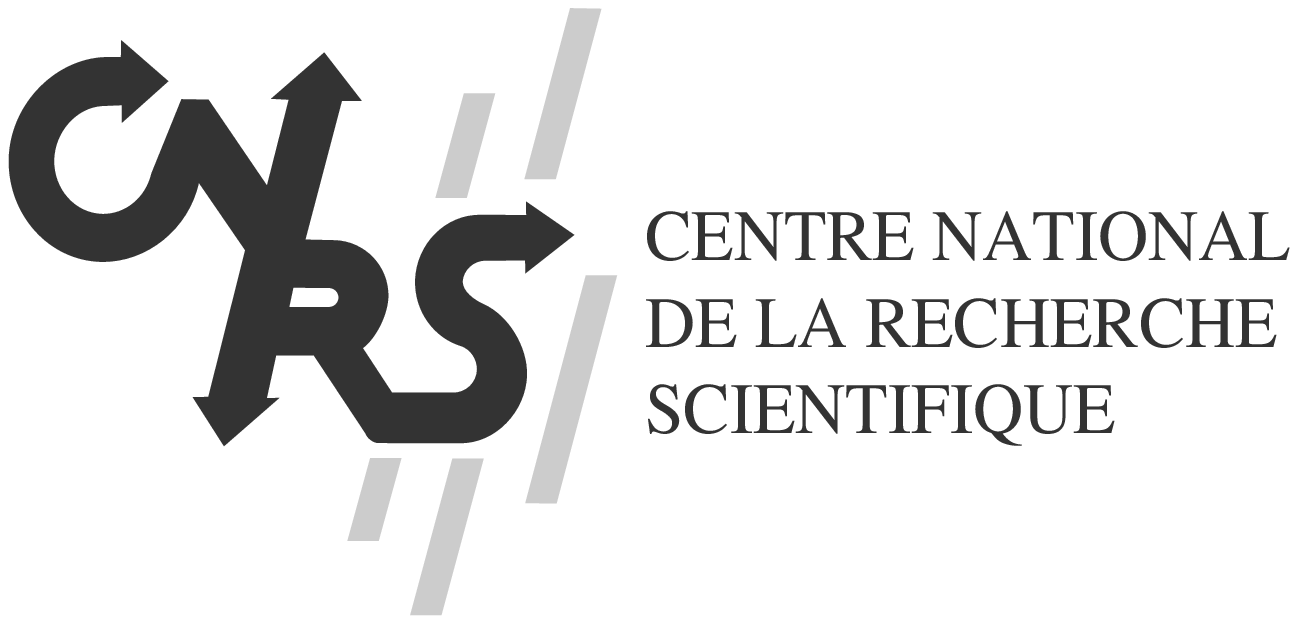}
\end{center}
\figskip
}
%

\end{titlepage}

\section{INTRODUCTION}
\label{section:introduction}

Binary systems of quasi-degenerate neutral mesons are undoubtedly among
the most interesting in particle physics, from both experimental and
theoretical points of view. It is in particular thanks to them that the
intriguing phenomenon of $CP$ violation \cite{ChristensenCroninFitchTurlay}
\cite{Lee} has been discovered.

Such systems are beautiful test grounds for Quantum Mechanics (QM) and, indeed,
most approaches to their peculiarities do not go beyond this level
\footnote{Ambiguities that appear in this treatment were recently outlined in
\cite{Machet}.}
; it is only recently that the need arose of a treatment in the
framework on Quantum Field Theory (QFT) \cite{BlasoneCapolupoRomeiVitiello}
\cite{Beuthe} (it was actually mainly motivated for the
leptonic sector after the discovery of neutrino oscillations).
However, conceptual problems still remained, in particular concerning the
existence of two different sets of mass eigenstates, belonging respectively
to the ``in'' and ``out'' spaces (see \cite{BrancoLavouraSilva}
\cite{AlvarezGaumeKounnasLolaPavlopoulos} and references therein).
General constraints cast by analyticity properties were never explicitly
written, and the ones cast by discrete symmetries often written with
conventions which forbade a full generality.
The formalism of a mass matrix also seemed never to be cast in doubt, though
its existence, as we shall see, can only be assumed in a certain
approximation.

All these open questions,  and the growing need for precise criteria to test
discrete symmetries, made necessary an exhaustive investigation of these
systems in QFT.  This is what we propose here.

The plan of the paper is the following: 

$\bullet$ In section \ref{section:constraints} we give the general definitions and
notations for the propagator of a binary system of neutral mesons, and
deduce on the most general ground the constraints cast on it by
analyticity, positivity, and the discrete symmetries $C$, $P$, $CP$ and
$TCP$. All arbitrary phases have been kept in the definition of discrete
symmetries, which make our formul\ae\, more general than the ones in
\cite{Sachs}; this has influence on particular on the role held by Lorentz
invariance in the deduction of the symmetry properties of the propagator.

In subsection \ref{subsub:massmat}, we show how the introduction of a mass
matrix can only be done in a linear approximation to the inverse
propagator. This casts restrictions on it, which will be made explicit
in subsection \ref{section:massmatrix}.

This section is completed by the long appendix \ref{section:disconst} which
explicitly  gives all demonstrations concerning the role of discrete
symmetries,  and provides a detailed discussion of the
special role of $TCP$. In particular, in the case of unstable
particles under concern which necessitates the introduction of a
non-hermitian Lagrangian, two ways of implementing the $TCP$ symmetry, that
we call the conventions of Wightman and of Schwinger-Pauli are discussed in
detail.

$\bullet$ Section \ref{section:matrices} is dedicated to the diagonalization of the
propagator, with a special emphasis on the property of normality.

\quad $\ast$ $CP$ invariance entails that the propagator is a special type
of normal matrix, and
subsection \ref{subsection:normalCP} deals with normal propagators and $CP$
eigenstates; we show that, if one wants furthermore to implement the
constraints set by $TCP$ invariance, the $CP$ violating parameter for a
general normal propagator is constrained to be purely imaginary, which is
in contradiction with experiments.

\quad $\ast$ subsection \ref{subsection:nonnormal} deals with non-normal
propagators.

We first recall two different ways of diagonalizing a complex mass matrix:
by a bi-unitary transformation and by using a bi-orthogonal basis. These
procedures are inequivalent, as will be explicitly shown in subsection
\ref{subsub:nobiunit}.

We then define, as they should be, the physical masses of the neutral
kaons, as the poles of their full propagator.

Next we explicitly diagonalize the $TCP$ invariant propagator by using
a bi-orthogonal basis and determine its physical (mass) eigenstates.
We determine all $CP$ violating parameters and show $TCP$ invariance does
not entail that the $CP$ parameter $\epsilon_L$ of $K_L$ is identical to
the one $\epsilon_S$ of $K_S$. We study their dependence on an arbitrary
rephasing of $K^0$ (and $\overline{K^0}$) and show that  their real and
imaginary parts depend on this phase;  physically relevant quantities are
of course, phase invariant.
This smooths out conceptual difficulties linked
to the existence of two sets of eigenstates, ``in'' and ``out''.
The study of the $CP$ violating parameters is completed in appendix
\ref{section:epsilon}.
We then give the explicit form of all propagating mass eigenstates
in terms of the $CP$ violating parameters.

We show the non-trivial way in which $TCP$ invariance is realized.
At each of the two scales $q^2 = M_{K_L}^2$ and $q^2 = M_{K_S}^2$,
the propagator has two sets of eigenstates: one corresponds to the
propagating $K_L$ (``in'' and ``out'') or $K_S$ (``in'' and ``out''),
and the other one does not propagate (we call it spurious).
At any given $q^2$, $TCP$ invariance needs the two sets of
eigenstates corresponding to this $q^2$, and, in particular, for $q^2$ equal
to any of the two physical masses, both the propagating and the spurious
states are essential.
Both sets of states are also needed for the completeness relation
at a given $q^2$.

We then show why bi-unitary transformations are not suitable to diagonalize
the propagator of the neutral mesons: while they yield the correct physical
masses and propagating eigenstates, their ``spurious'' eigenstates differ.

Last, we emphasize the role of the non-vanishing $\epsilon_L -
\epsilon_S$ by depicting the simplified picture that arises when the
two $CP$ violating parameters are assumed (like in $TCP$ invariant QM)
to be identical.

$\bullet$ Section \ref{section:massmatrix} deals with the eventual
introduction of a mass matrix, like commonly done in QM.

\quad $\ast$ First we recall the role of hermitian and normal mass matrices
in QM, in relation with neutral mesons.

\quad $\ast$ Then we show how a mass matrix in QFT cannot
give consistent results for the systems of neutral mesons and cannot
describe faithfully all its properties, in particular $TCP$ symmetry with
different $CP$ violating parameters for $K_L$ and $K_S$ as was shown to
occur. This yields restrictions on quantum mechanical treatments of such
systems which, nevertheless, can provide a satisfying description of $CP$
violation.

$\bullet$ Section \ref{section:exper} is dedicated to calculating three
semi-leptonic asymmetries. They are all unambiguously expressed in terms of
the $CP$ violating parameters $\epsilon_L^{in}$ and $\epsilon_S^{in}$ of
the mass eigenstates $|\ K_L>_{in}$ and $|\ K_S>_{in}$. We suppose that the
$\Delta S = \Delta Q$ rule is satisfied.

\quad $\ast$ We first calculate the asymmetry $A_T$ measured in the CPLEAR
experiment, and give a result which is
independent of an arbitrary rephasing of $K^0$ (and $\overline{K^0}$), unlike
the often quoted QM result $A_T \approx 4 \Re(\epsilon)$;

\quad $\ast$ We next calculate the semi-leptonic charge asymmetries
$\delta_L$ and $\delta_S$, and give, there too, a formula invariant
by the rephasing of $K^0$,
unlike a customary approximation often quoted in QM.

We show that $\delta_S - \delta_L$
measures the difference between the $CP$ violating parameters of
the two mass eigenstates; this does not characterize $TCP$ violation,
unlike in QM.

\quad $\ast$ We use the quark picture to find an estimate of $\epsilon_S -
\epsilon_L$ and find $\epsilon_S - \epsilon_L \approx \epsilon
\frac{m_L - m_S}{2m_K} \approx 10^{-17}$.

\quad $\ast$ Last, we calculate the so-called $TCP$ asymmetry $A_{TCP}$.
This is achieved by evaluating the Feynman diagrams obtained from a $TCP$
invariant Lagrangian. It is shown to vanish, and, even when $\epsilon_L
\not = \epsilon_S$, to be a test of $TCP$ invariance.

$\bullet$ Section \ref{section:conclusion} is a general conclusion.

The citations have been limited to a few number; it is impossible to quote
all the literature devoted to so rich systems of particles, and it is
fortunate that excellent textbooks are now available
\cite{Belusevic}\cite{BrancoLavouraSilva} \cite{BigiSanda}. We refer the
reader to these and all the references contained there.

Unless in subsection \ref{subsub:deltaeps} where estimates will be done
with the help
of the quark picture for mesons, we will work in the framework of
 a renormalizable local quantum field theory where the two neutral kaons
are represented by a complex operator and its hermitian conjugate.
We shall also consider that the
instability of neutral mesons does not break the one-to-one relationship
between fields and particles though, truly speaking, the only
asymptotic fields are electrons, neutrinos and photons.
Since we restrict ourselves to a binary system and, at the same time, want
to account for the breaking of discrete symmetries which can only be
observed through their decays, we have to allow for complex masses
\cite{JacobSachs} and a non-hermitian Lagrangian. The subtle interplay
with, in particular, $TCP$ symmetry is discussed in the Appendix
\ref{section:disconst}.

\section{THE PROPAGATOR; GENERAL CONSTRAINTS}
\label{section:constraints}

\subsection{DEFINITION AND NOTATIONS}
\label{subsection:definition}

Since we are dealing with complex (matricial) functions of a complex variable
$z$, it is essential to clearly set the notations and conventions which
will be used throughout this work.

If $z = x + iy, x,y \in {\mathbb R}$ is a complex number, its conjugate is
$\bar z = x - iy$; its real part is noted $\Re(z)$ and its imaginary part
$\Im(z)$.

If $f(z)$ is a complex function of the complex variable $z$, for example
$f(z) = a z^2 + bz + c, a,b,c \in {\mathbb C}$,  its complex conjugate
is noted $\bar f(\bar z)$ or $\overline{f(z)}$, and, for the example proposed, one has
$\bar f(\bar z) = \bar a \bar z^2 + \bar b \bar z + \bar c$.
The notation $\bar f(z) = \bar a z^2 + \bar b z + \bar c$ (for the given
example) can also be useful.

If $F(z) = \left( \begin{array}{cc} f_1(z) & f_2(z) \cr f_3(z) & f_4(z)
\end{array}\right)$ is a complex $2 \times 2$ matrix the elements of which
are complex
functions of the complex variable $z$, its hermitian conjugate is noted
$F^\dagger(\bar z) = \left( \begin{array}{cc} \overline{f_1}(\bar z) &
\overline{f_3}(\bar z) \cr \overline{f_2}(\bar z) & \overline{f_4}(\bar z)
\end{array}\right)$, and will also be noted $[F(z)]^\dagger$
\footnote{At any value of $z$, the elements of $F^\dagger(\bar z)$, which are
complex numbers, coincide with the elements of the hermitian conjugate
$[F(z)]^\dagger$ of $F(z)$ at the same value of $z$.}
.
In the case when $f_1$, $f_2$, $f_3$
and $f_4$ are polynomial functions of $z$, like in the example given above
for $f(z)$, it is also convenient to define $F^\dagger(z)$ which is
obtained from $F(z)$ by:\newline\null
- taking the transposed of $F(z)$;\newline\null
- changing all the coefficients of the $z$ monomials into their complex
  conjugates;\newline\null
- leaving $z$ unchanged.

The transposed of an operator $\cal O$ is noted ${\cal O}^T$ and its
 hermitian conjugate is noted ${\cal O}^\dagger$.

Unless specified, all propagators and mass matrices are written
in the $(K^0,\overline{K^0})$ basis.

Let $\varphi_{K^0}(x)$ be the Heisenberg operator for $K^0$ at space time point
$x = (\vec x,t)$ and $\varphi_{K^0}(\vec x)$ the corresponding
Schr\oe dinger operator (see also subsection \ref{subsub:CPanti}).
Since other fields will be related to it, we shall often omit the
corresponding subscript, writing instead $\varphi$ when it is the only one
appearing in a formula.

The Heisenberg field $\varphi_{\overline{K^0}}(\vec x,t)$ associated with
$\overline{K^0}$ is defined in terms of
$\varphi_{K^0}(\vec x,t)$ in subsection \ref{subsub:chacon} of
 appendix \ref{section:disconst};
this introduces two arbitrary phases $\alpha$ and $\delta$.
$C$ being the charge conjugation operator (operating on Schr\oe dinger
fields):
\begin{equation}
C \varphi_{K^0}(\vec x) C^{-1} =
                   e^{-i\alpha} \varphi_{\overline{K^0}}(\vec x),
\label{eq:cop}
\end{equation}
and
\begin{equation}
\varphi_{\overline{K^0}}(\vec x)= e^{-i\delta}\varphi^\dagger_{K^0}(\vec x).
\label{eq:barda}
\end{equation}
lead to
\begin{equation}
C \varphi_{K^0}(\vec x) C^{-1} =
                   e^{-i(\alpha+\delta)} \varphi^\dagger_{K^0}(\vec x).
\label{eq:cop2}
\end{equation}
In $x$ space, the Feynman propagator $\Delta(x)$ is a $2 \times 2$ matrix
function
which connects the $K^0$ and $\overline{K^0}$ states to themselves and to each other;
it is expressed in terms of vacuum expectation values of
time-ordered products $T\{\varphi(x)\varphi(y)\}$ of Heisenberg fields
\begin{equation}
\Delta(x) = \left(\begin{array}{rr}
                d(x) & -g(x) \cr
                -h(x) & f(x) \end{array}\right),
\label{eq:propx}
\end{equation}
with, using (\ref{eq:bardag})
\footnote{\cite{Sachs} uses different conventions; though we work with the
same type of approach, we found necessary to reinstall all phases that were
canceled there, to take out the parity operation from the definition of
$\overline{K^0}$ from $K^0$ and to come back to the basic definition of
antiparticles \cite{Stora}. This removes all ambiguities and fortuitous
coincidences in our demonstrations and results.}

\vbox{
\begin{eqnarray}
d(x)&=&<K^0\ |\ \Delta(x)\ |\ K^0> =
<0\ |\ T\{\varphi_{K^0}(\frac{\vec x}{2},\frac{t}{2})\
\varphi^\dagger_{K^0}(-\frac{\vec x}{2},-\frac{t}{2})\}\ |\ 0>,\cr
f(x)&=&<\overline{K^0}\ |\ \Delta(x)\ |\ \overline{K^0}> =
<0\ |\ T\{\varphi_{\overline{K^0}}(\frac{\vec x}{2},\frac{t}{2})\
        \varphi^\dagger_{\overline{K^0}}(-\frac{\vec x}{2},-\frac{t}{2})\}\ |\
0>\cr
&=& <0\ |\ T\{\varphi^\dagger_{{K^0}}(\frac{\vec x}{2},\frac{t}{2})\
        \varphi_{{K^0}}(-\frac{\vec x}{2},-\frac{t}{2})\}\ |\ 0>,\cr
-g(x)&=&<K^0\ |\ \Delta(x)\ |\ \overline{K^0}> =
<0\ |\ T\{\varphi_{K^0}(\frac{\vec x}{2},\frac{t}{2})\
        \varphi^\dagger_{\overline{K^0}}(-\frac{\vec x}{2},-\frac{t}{2})\}\ |\ 0>,\cr
&=& e^{i\delta}<0\ |\ T\{\varphi_{K^0}(\frac{\vec x}{2},\frac{t}{2})\
        \varphi_{{K^0}}(-\frac{\vec x}{2},-\frac{t}{2})\}\ |\ 0>,\cr
-h(x)&=&<\overline{K^0}\ |\ \Delta(x)\ |\ K^0> =
<0\ |\ T\{\varphi_{\overline{K^0}}(\frac{\vec x}{2},\frac{t}{2})\
        \varphi^\dagger_{K^0}(-\frac{\vec x}{2},-\frac{t}{2})\}\ |\ 0>\cr
&=& e^{-i\delta}<0\ |\ T\{\varphi^\dagger_{K^0}(\frac{\vec x}{2},\frac{t}{2})\
        \varphi^\dagger_{K^0}(-\frac{\vec x}{2},-\frac{t}{2})\}\ |\ 0>.
\label{eq:prop}
\end{eqnarray}
}
A theory is called ``Lorentz invariant'' when it is invariant by the proper
orthochronous Lorentz group $L^\uparrow_+$. Now, and this will be
important in the following, in particular in our discussion of $CP$
transformation, because we are here dealing with scalar fields,
$\Delta(\vec x,t)$ can only be a
function of $(|\vec x|^2-t^2)$, and, in particular, though space inversion,
which has determinant $(-1)$ is not part of $L^\uparrow_+$,
\begin{equation}
\Delta(\vec x,t)= \Delta(-\vec x,t).
\label{eq:Lorentz}
\end{equation}

\subsection{THE PROPAGATOR IN FOURIER SPACE; RENORMALIZATION}
\label{subsection:fourier}

In Fourier $(p^2=z)$ space, the matrix elements of $\Delta(z)$ are the
Fourier transformed of the ones of $\Delta(x)$ in (\ref{eq:prop})
\cite{Sachs}, and  we write
\begin{equation}
\Delta(z) = \left(\begin{array}{rr}
                d(z) & -g(z) \cr
                -h(z) & f(z) \end{array}\right).
\label{eq:propp}
\end{equation}
We assume from now on-wards that we operate in the framework of a
renormalizable quantum field theory for mesons.
$\varphi$ stands for the renormalized kaon field, and we note $\varphi_0$
the bare kaon field. All quantities occurring in (\ref{eq:prop}) and
(\ref{eq:propp}) are the renormalized ones.

$\varphi$ and $\varphi_0$ are connected as usual by
\begin{equation}
\varphi_0 = \sqrt{Z_{K^0}}\varphi,
\label{eq:renorphi}
\end{equation}
where $Z_{K^0}$ is the renormalization constant of the kaon field.

(\ref{eq:renorphi}) and the definitions (\ref{eq:prop}) entail that
$d(z)$, $f(z)$, $g(z)$ and $h(z)$ occurring in (\ref{eq:propp})
are connected to the bare  $d_0$, $f_0$, $g_0$ and $h_0$ defined below

\vbox{
\begin{eqnarray}
d_0(x) &\equiv&
   <0\ |T\{\varphi_0(\frac{\vec x}{2},\frac{t}{2})\
        \varphi_0^\dagger(-\frac{\vec x}{2},-\frac{t}{2})\}|\ 0>,\cr
f_0(x) &\equiv&
   <0\ |T\{\varphi_0^\dagger(\frac{\vec x}{2},\frac{t}{2})\
        \varphi_0(-\frac{\vec x}{2},-\frac{t}{2})\}|\ 0>,\cr
-g_0(x) &\equiv&
   e^{i\delta}<0\ |T\{\varphi_0(\frac{\vec x}{2},\frac{t}{2})\
        \varphi_0(-\frac{\vec x}{2},-\frac{t}{2})\}|\ 0>,\cr
-h_0(x) &\equiv&
   e^{-i\delta}<0\ |T\{\varphi_0^\dagger(\frac{\vec x}{2},\frac{t}{2})\
        \varphi_0^\dagger(-\frac{\vec x}{2},-\frac{t}{2})\}|\ 0>;
\label{eq:propbare}
\end{eqnarray}
}
 by
\begin{equation}
d(z) = \frac{1}{\sqrt{Z_{K^0}}(\sqrt{Z_{K^0}})^\dagger} d_0,\
f(z) = \frac{1}{(\sqrt{Z_{K^0}})^\dagger \sqrt{Z_{K^0}}} f_0,\
g(z) = \frac{1}{Z_{K^0}} g_0,\
h(z) = \frac{1}{Z_{K^0}^\dagger} h_0.
\label{eq:reno}
\end{equation}
(\ref{eq:propp}) yields the following renormalized inverse propagator
\begin{equation}
\Delta^{-1}(z) \equiv \frac{1}{d(z)f(z)-g(z)h(z)}
\left(\begin{array}{rr}
                                  f(z) & g(z) \cr
                                  h(z) & d(z) \end{array}\right)
\equiv \left(\begin{array}{rr}  a(z) &  -b(z) \cr
                          -c(z) &  d(z) \end{array}\right).
\label{eq:invprop}
\end{equation}
In (\ref{eq:propbare}) we have introduced the bare $\varphi_0$ and
$\varphi_0^\dagger$ fields and supposed that
\begin{equation}
C\varphi_0(\vec x)C^{-1} =
e^{-i(\alpha+\delta)}\varphi_0^\dagger(\vec x). \label{eq:CP0}
\end{equation}
indeed, in a renormalizable theory, the counterterms (unlike the
finite terms) are of the same nature as the operators present in
the initial Lagrangian \cite{Roman}, and they respect in particular the way in
which the fields transform by discrete symmetries and by complex
(hermitian) conjugation; so, if the bare kaons fields
are related to each other by charge conjugation in a certain
way,  the renormalized fields should be
related to each other in the same way; likewise, since the
counterterms control  the renormalization
  constants $Z_{K^0}$ and $Z_{\overline{K^0}}$,
the latter must satisfy (from (\ref{eq:barda}), (\ref{eq:CP0}) and
(\ref{eq:Cop2}))
\begin{equation}
Z_{\overline{K^0}} = Z_{K^0}^\dagger,
\label{eq:ZZ}
\end{equation}
such that, in all formul\ae,  $(\sqrt{Z_{K^0}})^\dagger$ can be
replaced with $\sqrt{Z_{\overline{K^0}}}$; furthermore, both are
calculated from the evaluation of Green functions in the
ultraviolet regime, that is far from any cut or physical
singularity, which entails that they must be real
 and, accordingly
\begin{equation}
Z_{\overline{K^0}} = Z_{K^0}.
 \label{eq:ZZ1}
\end{equation}

 From (\ref{eq:reno}) and (\ref{eq:invprop}), one gets the
following relations between the renormalized and bare components
of the inverse propagator

\vbox{
\begin{eqnarray}
&& a = {\sqrt{Z_{K^0}}(\sqrt{Z_{K^0}})^\dagger} \frac
{d_0}{d_0f_0-g_0 h_0}={\sqrt{Z_{K^0}}(\sqrt{Z_{K^0}})^\dagger} \hat a_0,\cr
&& d = {(\sqrt{Z_{K^0}})^\dagger \sqrt{Z_{K^0}}} \frac
{a_0}{d_0f_0-g_0 h_0}={\sqrt{Z_{K^0}}(\sqrt{Z_{K^0}})^\dagger} \hat d_0,\cr
&& b = {Z_{K^0}}\frac{-g_0}{d_0f_0-g_0 h_0}=
{Z_{K^0}}b_0,\cr
&& c = {Z_{K^0}^\dagger}\frac{-h_0}{d_0f_0-g_0 h_0}=
{Z_{K^0}^\dagger} c_0.
\label{eq:scal}
\end{eqnarray}
}

\subsection{ANALYTICITY AND POSITIVITY}
\label{subsection:anacons}

\subsubsection{K\"allen-Lehmann representation \cite{Stora}}
\label{subsub:KL}

The propagator can be demonstrated, with very general hypothesis
\footnote{Lorentz and translation invariance.}
, to satisfy a K\"allen-Lehmann representation, which writes, in Fourier space
\begin{equation}
\Delta(z) = \int_0^\infty d(k^2) \frac{\rho(k^2)}{k^2-z},
\label{eq:KaLe}
\end{equation}
where, eventually, $z$  gets close to the cut on the real axis by staying in
the physical upper half-plane ($z \rightarrow (p^2 + i\varepsilon), p^2 \in
{\mathbb R}$).

Since the propagator is a matrix, so is the spectral function, the elements
of which we shall call respectively $\rho_d$, $\rho_f$, $-\rho_g$,
$-\rho_h$. One has ($p_n$ being the momentum of the state $n$)

\vbox{
\begin{eqnarray}
\rho_d(k^2) &=&\sum_n <0\ |\varphi(\vec 0,0)|\ n><n\ | \varphi^\dagger(\vec
0,0)|\ 0>,\cr
\rho_f(k^2) &=&\sum_n <0\ |\varphi^\dagger(\vec 0,0)|\ n><n\ | \varphi(\vec
0,0)|\ 0>,\cr
-\rho_g(k^2) &=&e^{i\delta}\sum_n <0\ |\varphi(\vec 0,0)|\ n><n\ |
\varphi(\vec 0,0)|\ 0>,\cr
-\rho_h(k^2) &=&e^{-i\delta}\sum_n <0\ |\varphi^\dagger(\vec 0,0)|\ n><n\ |
\varphi^\dagger(\vec 0,0)|\ 0>,\cr
&& p_n^2 = k^2, p_n^0 > 0.
\label{eq:spectral}
\end{eqnarray}
}
Since $<0\ |\varphi(\vec 0,0)|\ n> = \overline{<n\ |\varphi^\dagger(\vec 0,0)|\ 0>}$, one
gets the constraints
\begin{eqnarray}
\rho_d(k^2) &=& \overline{\rho_d}(k^2) = \sum_n\left|<0\ |\varphi(\vec 0,0)|\
n>\right|^2 \geq 0,\cr
\rho_f(k^2) &=& \overline{\rho_f}(k^2) = \sum_n\left|<0\ |\varphi^\dagger(\vec
0,0)|\
n>\right|^2 \geq 0,\cr
\rho_g(k^2) &=& \overline{\rho_h}(k^2).
\label{eq:spancons}
\end{eqnarray}
The spectral function is accordingly a positive hermitian matrix
\footnote{This cannot be carelessly transposed to the propagator since, in
particular, $z$ in (\ref{eq:KaLe}) spans the complex plane.}
.

A consequence is that the propagator $\Delta(z)$  is an holomorphic
function in the complex $z$ plane outside the cuts
\footnote{In the case under concern, two cuts on the real axis start
respectively at $z= (2m_\pi)^2$ and $z = (3m_\pi)^2$.}
, which satisfies \cite{StreaterWightman}
\begin{equation}
\Delta(z) = [\Delta(\bar z)]^\dagger.
\label{eq:reality}
\end{equation}
Indeed, (\ref{eq:reality}) writes, using the hermiticity of $\rho$
\footnote{Since $k^2$ spans the real axis, $\rho(k^2)$ is a complex (matricial)
function of a real variable, and the notation $\rho^\dagger(k^2)$ is
unambiguous.}
.

\begin{equation}
\Delta(\bar z) = \int_0^\infty d(k^2) \frac{\rho(k^2)}{k^2-\bar z},
\quad
[\Delta(\bar z)]^\dagger = \int_0^\infty d(k^2)
\left[\frac{\rho(k^2)}{k^2-\bar z}\right]^\dagger =
\int_0^\infty d(k^2) \frac{\rho^\dagger(k^2)}{k^2-z}.
\label{eq:nota2}
\end{equation}

This general property should be distinguished from the (Schwarz)
reflection principle or its refined version called the ``edge of the wedge''
theorem \cite{StreaterWightman}; indeed, as soon as complex coupling
constants can enter the game, in particular to account for $CP$
violation, the discontinuity on the cut is no longer the sole origin for
the imaginary part of the propagator; it can be non-vanishing outside the
cut (as can be checked in a quark model), which is likely to invalidate
the principle of reflection.

\subsubsection{The linear approximation: introducing a mass matrix
\cite{Stora}}
\label{subsub:massmat}

We show here how a mass matrix can be introduced, which can describe
unstable particles and at the same time respect the positivity and
analyticity properties of the propagator, and how it can only be
considered as an approximation
\footnote{This section has been written thanks to the help of R. Stora.}.

The imaginary part of $\Delta(z)$
\begin{equation}
\Im(\Delta(z)) = \frac{\Delta(z) - (\Delta(z))^\dagger}{2i}=
\frac{\Delta(z) - \Delta^\dagger(\bar z)}{2i}
\end{equation}
rewrites, using (\ref{eq:KaLe}),
\begin{equation}
\Im (\Delta(z)) = \int_0^\infty d(k^2) \left( \frac{\rho(k^2)}{k^2-z}
- \frac{\rho^\dagger(k^2)}{k^2 -\bar z}\right);
\end{equation}
because of the constraints (\ref{eq:spancons}), one has for the four matrix
elements of $\Delta(z)$

\vbox{
\begin{eqnarray}
\Im(d(z)) &=& \int_0^\infty  \frac{(z-\bar
z)|\rho_d(k^2)|}{|k^2-z|^2},\cr
\Im(f(z)) &=& \int_0^\infty  \frac{(z-\bar
z)|\rho_f(k^2)|}{|k^2-z|^2},\cr
\Im(-g(z)) &=& \int_0^\infty \left( \frac{-\rho_g(k^2)}{k^2-z}
+ \frac{\overline{\rho_h}(k^2)}{k^2-\bar z}\right) =
\int_0^\infty \frac{-\rho_g(k^2)(z-\bar z)}{|k^2-z|^2},\cr
\Im(-h(z)) &=& \int_0^\infty \left( \frac{-\rho_h(k^2)}{k^2-z}
+ \frac{\overline{\rho_g}(k^2)}{k^2-\bar z}\right)
=\int_0^\infty \frac{-\rho_h(k^2)(z-\bar z)}{|k^2-z|^2},
\end{eqnarray}
}
such that, the imaginary part of the (matricial) Feynman propagator
is $(z-\bar z)$ times a positive hermitian matrix, and its sign is thus
 always the sign of $(z-\bar z)$.

If this property is true for the propagator $\Delta(z)$, it is also true
for its inverse $\Delta^{-1}(z)$. This is the property that we want to
preserve when approximating the inverse propagator.

Close to the poles, a linear approximation of $\Delta^{-1}$ should be
suitable,
\begin{equation}
\Delta^{-1}(z) \approx Az + B,
\end{equation}
such that
\begin{equation}
\Im(Az + B) =\frac {A-A^\dagger}{2i}\frac{z+\bar z}{2}
+ \frac{A + A^\dagger}{2}\frac{z-\bar z}{2i} + \frac{B-B^\dagger}{2i}.
\end{equation}
When $A=A^\dagger$ is a positive hermitian matrix, the sign of the first
two terms is indeed the same as the sign of $\Im(z)$.
The property of positivity is true everywhere only if $B=B^\dagger$;
in this case, the mass matrix is hermitian, its eigenvalues
are real and cannot describe unstable particles. However,
if one only wants to preserve this property in the upper (physical) half plane
$\Im(z) \geq 0$, it is enough to have $\Im(B) \geq 0$.
If this is so, then, writing $B= B_1 + i B_2, B_2 \geq 0$, one has
\begin{equation}
\Delta^{-1}(z) \approx \sqrt{A}\left( z + \frac{1}{\sqrt{A}} (B_1 + iB_2)
\frac{1}{\sqrt{A}}\right)\sqrt{A}
= \sqrt{A}\left(z - \left({m^{\{2\}}}-i\frac{{\Gamma^{\{2\}}}}{2}\right)\right)\sqrt{A};
\end{equation}
the mass matrix
\footnote{The superscript ``$\{2\}$''  which appear in $M^{\{2\}}$,
${m^{\{2\}}}$ and ${\Gamma^{\{2\}}}$
are to remind that these quantities have dimension $[mass]^2$.}
 is accordingly
\begin{equation}
{M^{\{2\}}}\equiv {m^{\{2\}}}-i\frac{{\Gamma^{\{2\}}}}{2}
= -\frac{1}{\sqrt{A}} (B_1 + iB_2)\frac{1}{\sqrt{A}},\text{with}\
{\Gamma^{\{2\}}} \geq 0, A=A^\dagger.
\label{eq:mama}
\end{equation}
It is no longer hermitian and can accordingly describe unstable kaons.

Since ${\Gamma^{\{2\}}} \geq 0$, the zeroes of the approximate inverse propagator
(poles of the approximate propagator) are located  in the lower
(unphysical) half plane.

The hermitian matrix $A$ normalizes the states.

\subsection{DISCRETE SYMMETRIES AND LORENTZ INVARIANCE}
\label{subsection:discons}

The first two paragraphs of this section summarizes the results obtained and
demonstrated at length in Appendix \ref{section:disconst} for the
propagator.

The next paragraphs demonstrate which constraints can be obtained on the
spectral function, using the two possible conventions for $TCP$
transformations, the one of Wightman and the one of Schwinger-Pauli.

\subsubsection{$\boldsymbol{CP}$ symmetry}
\label{subsub:CPcons}

$CP$ symmetry constrains the two diagonal elements of the propagator to be
identical, and the two antidiagonal elements to be related by
(\ref{eq:CPcons1}).
So, a $CP$ invariant kaon propagator is in particular a (special type of)
normal matrix; this leaves {\em a priori} for a general normal
propagator the possibility
to describe $CP$ violating theories.
We indeed investigate in subsection \ref{subsub:normalpropag}
the case of a general normal
propagator, and show that, then, the $CP$ violating parameter is
non-vanishing but always lies on the imaginary axis
\footnote{The phase of the $CP$ violating parameter is not an
observable \cite{RPP}; in particular, asymmetries are proportional
to the real part of the $CP$ violating parameter
(see subsection \ref{subsection:semilep} for QFT).
A purely imaginary $\epsilon_L$ can nevertheless be
considered  to violate $CP$ when it cannot be brought back to
$0$ by a (constant) rephasing of the neutral kaons, as  shown 
in subsection \ref{subsub:normalpropag}.
This is however incompatible with experiment.
When direct $CP$ violation is allowed,
one gets, by quantum mechanical arguments \cite{RPP} and considering that
$\epsilon_L = \epsilon_S$,
$\eta_{+-}= \epsilon_L + \epsilon' + i \Im(A_0)/\Re(A_0)$, where
$A_0$ is the amplitude for the decay of $K^0$ into two pions in the isospin
$0$ channel; $\Im m(\epsilon_L)$ and $\Im(A_0)$ depend on the choice
of phase for the neutral kaons; only the phase of the direct $CP$
violating parameter $\epsilon'$ is physically relevant. The phase of
$\epsilon = \epsilon_L + i \Im(A_0)/\Re(A_0)$
is measured to be close to $43.4$ degrees while the phase $\phi_{+-}$ of
$\eta_{+-}$ is measured to be close to $43.5$ degrees  \cite{RPP}
and these two phases theoretically coincide in superweak models
which do not allow for direct $CP$ violation ($\epsilon'=0$) \cite{Okun1}.
Suppose now that, as predicted from
a normal $TCP$ invariant propagator in our model,  $\epsilon_L$
is purely imaginary. Since we do not allow for direct $CP$ violation,
one expects, supposing that the relations obtained by QM arguments give
results close to the one of QFT,
$\eta_{+-} = \epsilon_L +i \Im(A_0)/\Re(A_0)$; then $\eta_{+-}$ should
also be purely imaginary, which is in conflict with experiment.
\label{footnote:pureim}}
.

\subsubsection{$\boldsymbol{TCP}$ symmetry}
\label{subsub:TCP}

$TCP$ symmetry constrains the two diagonal elements of the kaon propagator
to be identical, and yields no constraint on the antidiagonal elements.

Accordingly, a $TCP$ invariant propagator can be normal or not.

\subsubsection{Constraints on the spectral function \cite{Stora}}
\label{subsub:disspectral}

One makes use of the notations and conventions explained in Appendix
\ref{section:disconst}.

$\bullet$ Constraints from $TCP$ symmetry,
using the convention of Wightman (see subsection \ref{subsection:Wightman}).

One uses (\ref{eq:WitLee}) to express, in (\ref{eq:spectral}),
$\varphi^\dagger(\vec 0,0) = \Theta\varphi(\vec 0,0)\Theta^{-1}$ and,
reciprocally (using $\Theta=\Theta^{-1}$)
$\varphi(\vec 0,0) = \Theta^{-1}\varphi^\dagger(\vec 0,0)\Theta =
\Theta\varphi^\dagger(\vec 0,0)\Theta^{-1}$.
 
This yields
\begin{equation}
\rho_d(k^2) = \sum_n <0\ |\Theta \varphi^\dagger(\vec 0,0)\Theta^{-1}|\ n><n\
|\Theta\varphi(\vec 0,0)\Theta^{-1}|\ 0>.
\end{equation}
The vacuum is invariant by $TCP$, $|\ 0> = \Theta|\ 0>$, and one supposes
furthermore that the spectrum is also $TCP$ invariant
$\sum_n|\ n><n\ | = \sum_n |\ \Theta\; n><\Theta\; n\ |$, which yields
\begin{equation}
\rho_d(k^2) = \sum_n <\Theta\; 0\ |\Theta \varphi^\dagger(\vec
0,0)\Theta^{-1}|\ \Theta\; n><\Theta\; n\
|\Theta\varphi(\vec 0,0)\Theta^{-1}|\ \Theta\; 0>.
\end{equation}
One uses next the antiunitarity (\ref{eq:antiunit2}) of the $\Theta$ operator
to get
\begin{equation}
\rho_d(k^2) = \sum_n <n\ |\varphi(\vec 0,0)|\ 0><0\
|\varphi^\dagger(\vec 0,0)|\ n>= \rho_f(k^2).
\end{equation}
The same procedure applied to the antidiagonal elements of $\rho(k^2)$ only
yields tautologies (like for the propagator) and thus no constraints.

$\bullet$ Constraints from $TCP$ symmetry,
using the Schwinger-Pauli convention (see subsection \ref{subsection:SchPau}).

Reading from right to left instead of from left to right, one gets
\begin{equation}
\Theta \rho_d(k^2) = \rho_f(k^2),\quad \Theta (-\rho_g(k^2))= -\overline{\rho_h}(k^2).
\label{eq:spectra5}
\end{equation}

$\bullet$ Constraints arising from $CP$ symmetry.

Following the same lines as in subsections
\ref{subsub:CPdia} and \ref{subsub:CPanti}, if $CP$ invariance holds and if one
supposes that the spectrum is $CP$ invariant
($\sum_n|\ n><n\ | = \sum_n |\ CP\;n><CP\;n\ |$ one gets
\begin{equation}
CP\rho_d(k^2) = \rho_f(k^2),\quad CP(-\rho_g(k^2)) =
e^{-2i\alpha}(-\rho_h(k^2)).
\label{eq:CPspec}
\end{equation}
The $CP$ constraints on the spectral function are the same as the ones for
the propagator.

\section{NORMAL VERSUS NON-NORMAL  PROPAGATOR; DIAGONALIZATION}
\label{section:matrices}

We recall the definition of a normal matrix:
\begin{equation}
M\ \text{normal}\ \Leftrightarrow [M,M^\dagger]=0.
\label{eq:normaldef}
\end{equation}
Normality is a remarkable property of matrices: any matrix that commutes
with its hermitian conjugate can be diagonalized by a single unitary
transformation; its right and left eigenstates accordingly coincide;
furthermore, unlike hermitian matrices, it admits complex eigenvalues
\cite{Lowdin},
which makes it specially suited to describe unstable particles
\cite{JacobSachs}.

When $CP$ is conserved, we have shown that the propagator of neutral kaons
must be normal. This will provides us with the most general $CP$
eigenstates in the $(K^0,\overline{K^0})$ basis.

It is very tempting to have a normal propagator in any circumstance, since
right eigenstates and left eigenstates coincide; we will show that this is
impossible, since any normal matrix with equal
diagonal elements (a $TCP$ invariant propagator must have equal diagonal
elements) yields
eigenstates with purely imaginary indirect $CP$ violating parameters
$\epsilon_L$ and $\epsilon_S$. So, in particular on the cut(s), the
propagator is non-normal, and there exist different right and left
eigenstates. We
will demonstrate that the appropriate way of diagonalizing the
propagator is by using a bi-orthogonal basis, and that is is not
fully equivalent to a bi-unitary transformation, like the one currently used for
fermions. The
``propagating states'' are unambiguously determined to be
the states which correspond to the poles of the full renormalized propagator.
The  $CP$ violating parameters of any pair of left and
right propagating states can now be anywhere in the
complex plane but have equal modulus, which is the physically
relevant property. Conceptual problems linked with the non normality of the
propagator on the cut and the subsequent existence of right
and left eigenstates are thus wiped out.
We will give the explicit form of all eigenstates and
indirect $CP$ violating parameters.

The structure of the eigenstates of the full propagator will be
investigated in  details and will reveal  in particular
the subtle way $TCP$ symmetry is realized. We will
exhibit the important role of states which correspond to a vanishing
residue of the propagator (zero norm states), that we call spurious.

\subsection{NORMAL MATRICES AND $\boldsymbol{CP}$ EIGENSTATES}
\label{subsection:normalCP}

\subsubsection{$\boldsymbol{CP}$ eigenstates}
\label{subsub:CPeig}

$CP$ conserving propagators are special types of normal matrices with their
two diagonal elements identical and their two antidiagonal elements
related by (\ref{eq:CPcons1}). Accordingly, we consider
\begin{equation}
\Delta_{CP}(z) = \left( \begin{array}{cc}
                     d(z)     &     e^{-i\alpha}l(z) \cr
              e^{i\alpha}l(z) &    d(z)
\end{array}\right).
\label{eq:DelCP}
\end{equation}
The eigenvalues $\lambda^{CP}_\pm(z)$ are the two solutions of the
characteristic equation of $\Delta_{CP}(z)$
\begin{equation}
\lambda^{CP}_\pm(z) = d(z) \pm l(z),
\label{eq:eigenCP}
\end{equation}
and the corresponding eigenvectors that we note $\left( \begin{array}{c}
u^{CP}_\pm(z) \cr v^{CP}_\pm(z)\end{array}\right)$ satisfy
\begin{equation}
r^{CP}_+(z) \equiv\frac{v^{CP}_+(z)}{u^{CP}_+(z)} = e^{i\alpha},\quad
r^{CP}_-(z) \equiv\frac{v^{CP}_-(z)}{u^{CP}_-(z)} = -e^{i\alpha},
\label{eq:eigenCP2}
\end{equation}
which are quantities independent of $z$: the $CP$ eigenstates, which are,
of course, function of the arbitrary phase $\alpha$ introduced in
(\ref{eq:cop}), do not change with $p^2$; this is why we call them directly
$K^0_1$ and $K^0_2$, explicitly:
\begin{equation}
|\ K^0_1> = \frac{1}{\sqrt{2}}( |\ K^0> + e^{i\alpha} |\ \overline{K^0}>),\quad
|\ K^0_2> = \frac{1}{\sqrt{2}}( |\ K^0> - e^{i\alpha} |\ \overline{K^0}>).
\label{eq:eigenCP3}
\end{equation}

\subsubsection{Normal propagators}
\label{subsub:normalpropag}

Let us now consider a general  normal propagator
\begin{equation}
\Delta_N(z) = \left( \begin{array}{cc}
                    d(z)    &      -g(z) \cr
                  -h(z)     &       f(z)  \end{array}\right),
\text{with}\ |g| = |h|\ \text{and}\ \bar h(d-f) = g(\bar d - \bar f).
\label{eq:Deltanorm}
\end{equation}
The condition at the right of (\ref{eq:Deltanorm}) are the condition for
the normality $[\Delta_N,\Delta_N^\dagger]=0$ of $\Delta_N$.

We introduce the phases $\alpha_g$ and $\alpha_h$ of $-g$ and $-h$ and the
conditions of normality become

\vbox{
\begin{eqnarray}
&&-g(z) = \rho(z)e^{i\alpha_g(z)}, -h(z)= \rho(z)e^{i\alpha_h(z)},\
\rho(z) \in {\mathbb R},\cr
&& (d(z) - f(z))- e^{i(\alpha_g(z)+\alpha_h(z))}(\bar d(z) -\bar f(z))=0
\Leftrightarrow d(z) - f(z) = |d(z) - f(z)|
e^{i\frac{\alpha_g(z)+\alpha_h(z)}{2}}.\cr
&&
\label{eq:normalcond}
\end{eqnarray}
}
It is convenient to introduce the following notations
\begin{eqnarray}
\sigma(z) &=& \frac{|d(z)-f(z)|}{2\rho(z)} = \frac{d(z) - f(z)}{2\rho(z)
e^{i\Sigma\alpha(z)}},\cr
\Sigma\alpha(z)&=& \frac{1}{2}(\alpha_g(z) + \alpha_h(z)),\cr
\Delta\alpha(z) &=& \frac{1}{2}(\alpha_g(z) - \alpha_h(z)).
\label{noteN}
\end{eqnarray}
The eigenvalues $\lambda^N_\pm(z)$ are given by
\begin{equation}
\lambda^N_\pm
=\frac{d(z)+f(z)}{2}\pm \rho(z) e^{i\Sigma\alpha(z)}\sqrt{1+\sigma^2(z)}.
\label{eq:rootsN}
\end{equation}
Writing the eigenvectors $\left(\begin{array}{c} u^N_\pm(z) \cr v^N_\pm(z)
\end{array}\right)$, and defining $r^N_\pm(z) = \frac{v^N_\pm(z)}
{u^N_\pm(z)}$ one gets
\begin{equation}
r^N_\pm(z)
=e^{i\Delta\alpha(z)} \left(\sigma(z) \pm \sqrt{1+\sigma^2(z)}\right).
\label{eq:rN}
\end{equation}
To determine the values of the indirect $CP$ violating parameter
$\epsilon^N(z)$, one goes to the basis of $CP$ eigenstates defined in
subsection \ref{subsub:CPeig} above. This gives
\footnote{For the eigenstates will subscript ``$+$'', $\epsilon_+$ is
defined as the ratio of the $K^0_2$ component over the $K^0_1$ component,
and for the eigenstate with subscript ``$-$'', $\epsilon_-$ is defined
 as the ration of the $K^0_1$
component over the $K^0_2$ component. So doing, we will match in the
following the usual
definitions of $\epsilon_L$ (for ``$+$'' states)
 and $\epsilon_S$ (for ``$-$'' states) for $K_L$ and $K_S$ mesons.}
\begin{eqnarray}
\epsilon_+^N(z) &=&
\frac{1-e^{-i\alpha}r^N_+(z)}{1+e^{-i\alpha}r^N_+(z)}
=
\frac{1-e^{i(-\alpha+\Delta\alpha(z))}\left(\sigma(z)
+ \sqrt{1+\sigma^2(z)}\right)}
{1+e^{i(-\alpha+\Delta\alpha(z))}\left(\sigma(z)
+ \sqrt{1+\sigma^2(z)}\right)},\cr
\epsilon_-^N(z) &=&
\frac{1+e^{-i\alpha}r^N_+(z)}{1-e^{-i\alpha}r^N_+(z)}
=
\frac
{1+e^{i(-\alpha+\Delta\alpha(z))}\left(\sigma(z) + \sqrt{1+\sigma^2(z)}\right)}
{1-e^{i(-\alpha+\Delta\alpha(z))}\left(\sigma(z) +
\sqrt{1+\sigma^2(z)}\right)},
\label{eq:epsilonN}
\end{eqnarray}
which is always purely imaginary when $d(z)=f(z)$, {\em i.e.} when $TCP$ is
satisfied, since this entails $\sigma(z)=0$ and
\begin{equation}
\epsilon_+^N(z)
\stackrel{TCP}{=} \frac{1- e^{i(-\alpha+\Delta\alpha(z))}}
{1+ e^{i(-\alpha+\Delta\alpha(z))}},\quad
\epsilon_-^N(z)
\stackrel{TCP}{=} \frac{1+ e^{i(-\alpha+\Delta\alpha(z))}}
{1- e^{i(-\alpha+\Delta\alpha(z))}}.
\label{eq:pureim}
\end{equation}
so, if $TCP$ is satisfied, for any value of $z=p^2$ where the propagator
is normal, its eigenstates have an indirect $CP$ violating parameter which
is purely imaginary; it cannot be brought back to $0$ by a constant
rephasing of the neutral kaons equivalent to choosing $\alpha=0$ since
the difference of phases $\Delta\alpha(z)$ between the antidiagonal
elements of the propagator, which depend on $z=q^2$,
also enters (\ref{eq:pureim}).

Normality of the propagator is consequently excluded;
indeed, as will be emphasized in section \ref{section:exper} (see footnote
\ref{footnote:pureim}),
a purely imaginary $CP$ violating parameter $\epsilon_L$ is incompatible
with experiment.

The other way to get $\epsilon$ non purely imaginary for a normal
propagator would be to keep $\sigma(z) \not= 0$, that is to abandon the
criteria of $TCP$ invariance; this is certainly not desired.

\subsection{NON-NORMAL MATRICES AND PROPAGATORS}
\label{subsection:nonnormal}

When studying kaon decays, one has to deal with a non-normal propagator.

\subsubsection{Diagonalization}
\label{subsub:diago}

The diagonalization of a non-normal complex matrix is not unique, and this
is why time has to be spent on this question
\footnote{The work \cite{Silva} is instructive, which emphasizes, in the
framework of QM, the importance of using  a
``reciprocal'' basis for the diagonalization of a non-normal effective mass
matrix.\label{footnote:Silva}}
.

$\bullet$\quad The first way to diagonalize a complex matrix is via a bi-unitary
transformation, that is two different unitary transformations, respectively
acting on the left and on the right; this is always how the quark mass
matrices are diagonalized
\footnote{When one chirality of fermions does not participate in
non-abelian weak interactions, like right-handed
fermions in the Glashow-Salam-Weinberg model, one can use a single unitary
transformation \cite{MachetPetcov}.}
.
Any given complex matrix $C(z)$ can always be diagonalized by two unitary
matrices $U(z)$ and $V(z)$ such that $(U(z))^\dagger C(z) V(z)=
diag(\mu_1(z), \mu_2(z))$; $U$ and $V$ respectively
diagonalize $C(z) (C(z))^\dagger$ and $(C(z))^\dagger C(z)$ (now $C$ and
$C^\dagger$ are supposed not to commute), and each of these two products
are hermitian and have real positive eigenvalues; $\mu_1$ and $\mu_2$ can
always be also chosen real and positive.
The eigenvalues of $C(z)$ determined in this way are not the roots of its
characteristic equation; instead, the square of these eigenvalues are the
roots of the characteristic equation of $CC^\dagger$ or $C^\dagger C$;
this leads to a different result, though, as we shall see, the poles
coincide.

$\bullet$\quad The second way to diagonalize a general complex mass matrix is by the
standard procedure of determining its eigenvalues as the roots
$\lambda_\pm(z)$ of its characteristic equation
\footnote{The right and left eigenvalues always coincide.}
, and then determining the
right and left eigenstates, respectively \hbox{$|\ R_\pm(z)>$} $ =
\frac{1}{n^R_\pm}\left(\begin{array}{c}
u_\pm(z) \cr v_\pm(z)\end{array}\right) \equiv \frac{1}{n^R_\pm}\left(u_\pm(z)
|\ K^0> + v_\pm^R |\ \overline{K^0}>\right)$ and
\hbox{$<L_\pm(z)\ |$} $ =
\frac{1}{n^L_\pm}\left( x_\pm(z),\   y_\pm(z)\right)
\equiv \frac{1}{n^L_\pm}\left(x_\pm(z) <K^0\ | + v_\pm^R <\overline{K^0}\ |\right)$
\footnote{There is no distinction between ``in'' and ``out'' states for
the flavor eigenstates $K^0$ and $\overline{K^0}$
(see for example the discussion in the third section of  \cite{Machet}).
One has $<K^0\ | = (|\ K^0>)^\dagger, <\overline{K^0}\ | = (|\
\overline{K^0}>)^\dagger$, $<K^0\ |\ K^0> = 1 = <\overline{K^0}\ |\
\overline{K^0}>$,
$<K^0\ |\ \overline{K^0}>=0=<\overline{K^0}\ |\ K^0>$.}
by the equations $C(z) |\ R_\pm(z)> = \lambda_\pm(z) |\ R_\pm(z)>$ and
$<L_\pm(z)\ | C(z) = <L_\pm(z)\ | \lambda_\pm(z)$. The normalization
conditions are then written between the two spaces of ``in'' (left) and ``out''
(right) states
\begin{eqnarray}
<L_+(z)\ |\  R_+(z)> = &1& = <L_-(z)\ |\  R_-(z)> ,\cr
<L_-(z)\ |\  R_+(z)> = &0& = <L_+(z)\ |\  R_-(z)>,
\label{eq:normalis}
\end{eqnarray}
which determines the normalization coefficients $n^R_\pm(z)$ and $n^L_\pm(z)$.
In general

\vbox{
\begin{eqnarray}
<R_\pm(z)\ |\  R_\pm(z)> \not= 1, \quad <L_\pm(z)\ |\  L_\pm(z)> \not= 1 ,\cr
<R_\mp(z)\ |\  R_\pm(z)> \not= 0,\quad  <L_\mp(z)\ |\  L_\pm(z)> \not=0,
\label{eq:normalis2}
\end{eqnarray}
}
where, for any vector, $< \ | = |\ >^\dagger$: the ``in'' eigenvectors do
not form a basis, nor the ``out'' eigenstates.

When dealing with constant matrices, the two procedures are different
and non equivalent. The second
procedure allows in particular complex eigenvalues, which is necessary for
a mass matrix of unstable states. However, we will see in subsection
\ref{subsub:nobiunit} that, when dealing with the full renormalized
propagator (with depend on $p^2$, while the two procedures
select the same physical masses and propagating
eigenstates, they however differ as far as  spurious
states are concerned; the latter play an essential role in the realization of
discrete symmetries, in particular $TCP$.

Discriminating the two procedures by the reality or not of their
eigenvalues is only valid for constant mass matrices; for $p^2$ dependent
propagators, this does not provide a criterion for rejecting biunitary
transformations.

\subsubsection{Physical masses}
\label{subsub:physmass}

In QFT, the physical masses are the values of $z=p^2$
which are poles of the the full renormalized propagator; accordingly, they
are determined by the equation
\begin{equation}
\text{At}\ z=\text{physical mass}\ ,\det(\frac{1}{\Delta(z)})=0.
\label{eq:physmass}
\end{equation}
We shall assume hereafter that there exist only two solutions to this
equation, $z_1 = M_L^2$ and $z_2 = M_S^2$; they are both complex numbers.

For the sake of convenience, we shall work in the following with the
$TCP$ invariant inverse propagator (see (\ref{eq:invprop}))
\begin{equation}
{\Delta}^{-1}(z) = \left(\begin{array}{cc}
                          a(z)    &    -b(z) \cr
                          -c(z)    &    a(z) \end{array}\right)
\label{eq:invpro}
\end{equation}
where
\begin{equation}
a(z) = \frac{f(z)}{f^2(z)-g(z)h(z)},\quad b(z) =
-\frac{g(z)}{f^2(z)-g(z)h(z)},\quad c(z) = -\frac{h(z)}{f^2(z)-g(z)h(z)}.
\label{eq:abc}
\end{equation}
Eigenstates of $\Delta^{-1}$ are of course the same of the ones of
$\Delta$, and the eigenvalues of the former are the inverse of the ones of
the latter.

The physical masses are accordingly defined by
\begin{equation}
a^2(z) = b(z)c(z),
\label{eq:physmass2}
\end{equation}
and we will choose, by convention
\begin{equation}
a(z_1) = - \sqrt{b(z_1)c(z_1)},\quad a(z_2) = + \sqrt{b(z_2)c(z_2)}.
\label{eq:physmass3}
\end{equation}

\subsubsection{$\boldsymbol{TCP}$ eigenstates}
\label{subsub:TCPeig}

At any given $z=p^2$, $\Delta^{-1}(z)$ has two eigenvalues $\lambda_+(z)$
and $\lambda_-(z)$
\begin{equation}
\lambda_+(z) = a(z) + \sqrt{b(z)c(z)}, \quad
\lambda_-(z) = a(z) - \sqrt{b(z)c(z)}.
\label{eq:ev1}
\end{equation}
To each of them corresponds one right eigenstate $|\ R_\pm(z)>_{in}$ and one
left eigenstate $_{out}<L_\pm(z)\ |$
\footnote{Since it will be used in subsection \ref{subsub:CPLEAR}, we give
here the explicit form of the eigenstates at any $z=q^2$.
\begin{eqnarray}
|\ R_+(z)>_{in} &=& \frac{1}{n(z)}\left(\sqrt{b(z)}|\ K^0> - \sqrt{c(z)}|\
\overline{K^0}>\right),\cr
|\ R_-(z)>_{in} &=& \frac{1}{n(z)}\left(\sqrt{b(z)}|\ K^0> + \sqrt{c(z)}|\
\overline{K^0}>\right),\cr
_{out}\!<L_+(z)\ | &=& \frac{1}{n(z)}\left(\sqrt{c(z)}<K^0\ | -
\sqrt{b(z)}<\overline{K^0}\ |\right),\cr
_{out}\!<L_-(z)\ | &=& \frac{1}{n(z)}\left(\sqrt{c(z)}<K^0\ | +
\sqrt{b(z)}<\overline{K^0}\ |\right),\cr
n^2(z) &=& 2\sqrt{b(z)c(z)}.
\label{eq:geneig}
\end{eqnarray}
}
; this occurs in particular at the two
physical masses $z=z_1$ and $z=z_2$, such that we have to deal with a total
of eight eigenvectors of $\Delta^{-1}(z)$, which will all be important, for
various reasons. They will be called
\begin{eqnarray}
&&|\ R_+(z_1)>_{in} = |\ K_L>_{in},\quad
|\ R_-(z_2)>_{in} = |\ K_S>_{in}, \cr
&& \cr
&&_{out}\!<L_+(z_1)\ | =  _{out}\!<K_L\ |,\quad
_{out}\!<L_-(z_2)\ | =  _{out}\!<K_S\ |,\cr
&& \cr
&&|\ R_+(z_2)>_{in} = |\ \tilde K_S>_{in},\quad
|\ R_-(z_1)>_{in} = |\ \tilde K_L>_{in}, \cr
&& \cr
&&_{out}\!<L_+(z_2)\ | =  _{out}\!<\tilde K_S\ |,\quad
_{out}\!<L_-(z_1)\ | =  _{out}\!<\tilde K_L\ |.
\label{eq:ev2}
\end{eqnarray}
As emphasized before
\begin{eqnarray}
&&_{out}\!<K_L\ | \not = (|\ K_L>_{in})^\dagger,\quad
_{out}\!<K_S\ | \not = (|\ K_S>_{in})^\dagger,\cr
&&_{out}\!<\tilde{K_L}\ | \not = (|\ \tilde{K_L}>_{in})^\dagger,\quad
_{out}\!<\tilde{K_S}\ | \not = (|\ \tilde{K_S}>_{in})^\dagger.
\end{eqnarray}
The first four eigenstates of (\ref{eq:ev2}) all have in common to
correspond to a {\em vanishing} eigenvalue of $\Delta^{-1}(z_1)$ or
$\Delta^{-1}(z_2)$, and the
last four to a non-vanishing eigenvalue; indeed, one
has, in virtue of (\ref{eq:ev1}) and (\ref{eq:physmass3})
\begin{eqnarray}
\lambda_+(z_1) = &0& = \lambda_-(z_2),\cr
\lambda_+(z_2) = 2a(z_2) = 2\sqrt{b(z_2)c(z_2)},&\ &
\lambda_-(z_1) = 2a(z_1) = -2 \sqrt{b(z_1)c(z_1)}.
\label{eq:ev3}
\end{eqnarray}
Only the first four eigenstates of (\ref{eq:ev3}) are propagating
eigenstates, and they correspond to the physical $K_L$ and $K_S$ mesons; we
shall study below the difference between their ``in'' and ``out''
states. The four other eigenstates are non propagating in the sense that
the corresponding  residues of the propagators at respectively
$\lambda_+(z_2)$ and $\lambda_-(z_1)$ are vanishing as can be easily
checked by making an expansion of the propagator for $z_2 \approx z \approx
z_1$; these states are zero norm states that we call ``spurious''.

They are however important and should not be neglected; we shall come back
at length on this point in subsection \ref{subsub:realTCP} and
\ref{section:exper} dealing with kaon decays,
but the theoretical argument is the following: at any $z$, the completeness
relation writes
\footnote{It is important to stress that the completeness relation cannot
involve both propagating states and that, in particular
$|\ K_L>_{in}\;_{out}\!<K_L\ | + |\ K_S>_{in}\;_{out}\!<K_S\ | \not = 1$.}
\begin{equation}
1 = |\ R_+(z)>\!{_{in}}\ {_{out}}\!<L_+(z)\ | + |\
R_-(z)>\!{_{in}}\ {_{out}}\!<L_-(z)\ |,
\label{eq:completeness}
\end{equation}
and this should stay in particular true at the physical poles $z=z_1$ and
$z=z_2$, in which case one of the two states appearing in the completeness
relation becomes a spurious state: the space of eigenvectors shrinks
to a one-dimensional space at the pole, and the propagator becomes a matrix
of rank 1.

Note furthermore that the propagator writes
\begin{equation}
\Delta(z) =
|\ R_+(z)>\!{_{in}}\frac{1}{\lambda_+(z)}{_{out}}\!<L_+(z)\ |
+ |\ R_-(z)>\!{_{in}}\frac{1}{\lambda_-(z)}{_{out}}\!<L_-(z)\ |,
\label{eq:pro2}
\end{equation}
which selects only the propagating state at each pole, but that the inverse
propagator (that is the quadratic renormalized Lagrangian) writes
\begin{equation}
\Delta^{-1}(z) =
|\ R_+(z)>\!{_{in}}{\lambda_+(z)}{_{out}}\!<L_+(z)\ |
+ |\ R_-(z)>\!{_{in}}{\lambda_-(z)}{_{out}}\!<L_-(z)\ |,
\label{eq:invpro2}
\end{equation}
which instead selects at each physical mass the spurious state.

The orthogonality relations that the eigenstates satisfy, which enable to
fix their normalization, are the following:

\vbox{
\begin{eqnarray}
&&\hskip 2cm _{out}\!<K_L\ |\ K_L>\!_{in} = 1,\cr
&&\hskip 2cm _{out}\!<K_S\ |\ K_S>\!_{in} = 1,\cr
&&\hskip 2cm _{out}\!<\tilde{K_L}\ |\ \tilde{K_L}>\!_{in} = 1,\cr
&&\hskip 2cm _{out}\!<\tilde{K_S}\ |\ \tilde{K_S}>\!_{in} = 1,\cr
&&{_{out}\!<\tilde K_L\ |\ K_L>\!_{in}} = 0 = {_{out}\!<K_L\ |\
\tilde K_L>\!_{in}},\cr
&&{_{out}\!<\tilde K_S\ |\  K_S>\!_{in}} = 0 = {_{out}\!< K_S\ |\
\tilde K_S>\!_{in}}.
\label{eq:norm3}
\end{eqnarray}
}

We now explicitly list all eigenstates of a $TCP$ invariant propagator
\footnote{
A remark is due here concerning the normalization of states in (\ref{eq:ev4}).
(\ref{eq:norm3}) allows the multiplication of a given $|\ >_{in}$ state by
a constant $1/N$ while the corresponding ${_{out}\!<\ |}$ state is
multiplied by $N$. Let us show that $N$ can only be a phase.
Using the time evolution induced by the Schr\oe dinger equation for
unstable particles (see subsection \ref{subsub:ATCP}), one gets then, for
example for the $K_L$ meson:
\begin{eqnarray}
|\ K_L(t)>_{in} &=& \frac{1}{N_L} e^{(im_L + \frac{\Gamma_L}{2})t}
 \frac{i}{\sqrt{2a(z_1)}}\left(
\sqrt{b(z_1)}|\ K^0> - \sqrt{c(z_1)}|\ \overline{K^0}> \right),\cr
_{out}\!<K_L(t)\ | &=& N_L e^{(-im_L - \frac{\Gamma_L}{2})t}
 \frac{i}{\sqrt{2a(z_1)}}
\left( \sqrt{c(z_1)}<K^0\ | - \sqrt{b(z_1)}<\overline{K^0}\ | \right),
\label{eq:teig}
\end{eqnarray}
where the mass of $K_L$ has been written $M_L = m_L -i\frac{\Gamma_L}{2}$.

(\ref{eq:teig}) yields in particular
\begin{eqnarray}
{_{in}\!<K_L(t)\ |\ K_L(t)>_{in}}&=& \frac{1}{|N_L|^2}e^{\Gamma_Lt}
\frac{1}{2|a(z_1)|}\left(|b(z_1)|+|c(z_1)|\right),\cr
{_{out}\!<K_L(t)\ |\ K_L(t)>_{out}}&=& {|N_L|^2}e^{-\Gamma_Lt}
\frac{1}{2|a(z_1)|}\left(|b(z_1)|+|c(z_1)|\right).
\label{eq:teig2}
\end{eqnarray}
However (see for example \cite{AlvarezGaumeKounnasLolaPavlopoulos}),
${_{out}\!<K_L(t)\ |\ K_L(t)>_{out}}$ is the time-reversed of
${_{in}\!<K_L(t)\ |\ K_L(t)>_{in}}$, such that $N_L$ must satisfy
\begin{equation}
\frac{1}{|N_L|^2} = T |N_L|^2 \Rightarrow |N_L|^2 = 1.
\end{equation}
}:

\vbox{
\begin{eqnarray}
|\ K_L>\!_{in} &=& \frac{i}{\sqrt{2a(z_1)}} \left(
\sqrt{b(z_1)}|\ K^0> - \sqrt{c(z_1)}|\ \overline{K^0}> \right)
=\frac{1}{n_L^{in}}\left(|\ K^0_2>+ \epsilon_L^{in}|\
K^0_1>\right),\cr
|\ K_S>\!_{in} &=&\frac{1}{\sqrt{2a(z_2)}}
\left( \sqrt{b(z_2)}|\ K^0> + \sqrt{c(z_2)}|\ \overline{K^0}> \right)
=\frac{1}{n_S^{in}}\left(\epsilon_S^{in}|\ K^0_2>+ |\ K^0_1>\right),\cr
|\ \tilde K_L>\!_{in} &=& \frac{i}{\sqrt{2a(z_1)}}
\left( \sqrt{b(z_1)}|\ K^0> + \sqrt{c(z_1)}|\ \overline{K^0}> \right)
=\frac{1}{n_{\tilde L}^{in}}
\left(|\ K^0_2>+ \epsilon_{\tilde L}^{in}|\ K^0_1>\right),\cr
|\ \tilde K_S>\!_{in} &=& \frac{1}{\sqrt{2a(z_2)}}
\left( \sqrt{b(z_2)}|\ K^0> - \sqrt{c(z_2)}|\ \overline{K^0}> \right)
=\frac{1}{n_{\tilde S}^{in}}
\left(\epsilon_{\tilde S}^{in}|\ K^0_2>+ |\ K^0_1>\right),\cr
_{out}\!<K_L\ | &=& \frac{i}{\sqrt{2a(z_1)}}
\left( \sqrt{c(z_1)}<K^0\ | - \sqrt{b(z_1)}<\overline{K^0}\ | \right)
=\frac{1}{n_L^{out}}\left(<K^0_2\ |+ \overline{\epsilon_L^{out}}<K^0_1\ |\right),\cr
_{out}\!<K_S\ | &=& \frac{1}{\sqrt{2a(z_2)}}
\left( \sqrt{c(z_2)}<K^0\ | + \sqrt{b(z_2)}<\overline{K^0}\ | \right)
=\frac{1}{n_S^{out}}\left(\overline{\epsilon_S^{out}}<K^0_2\ |+ <K^0_1\ |\right),\cr
_{out}\!<\tilde K_L\ | &=& \frac{i}{\sqrt{2a(z_1)}}
\left( \sqrt{c(z_1)}<K^0\ | + \sqrt{b(z_1)}<\overline{K^0}\ | \right)
=\frac{1}{n_{\tilde L}^{out}}
\left(<K^0_2\ |+ \overline{\epsilon_{\tilde L}^{out}}<K^0_1\ |\right),\cr
_{out}\!<\tilde K_S\ | &=& \frac{1}{\sqrt{2a(z_2)}}
\left( \sqrt{c(z_2)}<K^0\ | - \sqrt{b(z_2)}<\overline{K^0}\ | \right)
=\frac{1}{n_{\tilde S}^{out}}
\left(\overline{\epsilon_{\tilde S}^{out}}<K^0_2\ |+ <K^0_1\ |\right),\cr
&&
\label{eq:ev4}
\end{eqnarray}
}
where $K^0_1$ and $K^0_2$ can be found in (\ref{eq:eigenCP3}) and
where one can always take
\begin{eqnarray}
&&n_L^{in}=n_L^{out} = \sqrt{1+\epsilon_L^{in}\overline{\epsilon_L^{out}}},\quad
n_S^{in}=n_S^{out} = \sqrt{1+\epsilon_S^{in}\overline{\epsilon_S^{out}}},\cr
&&n_{\tilde L}^{in}=n_{\tilde L}^{out} =
          \sqrt{1+\epsilon_{\tilde L}^{in}\overline{\epsilon_{\tilde L}^{out}}},\quad
n_{\tilde S}^{in}=n_{\tilde S}^{out} =
        \sqrt{1+\epsilon_{\tilde S}^{in}\overline{\epsilon_{\tilde S}^{out}}}.
\label{eq:normeig}
\end{eqnarray}
Note that we have no {\em a priori} relations between states corresponding
to different values of $z$
\footnote{In particular, ${_{out}\!<K_S\ |\ K_L>_{in}} \not = 0$ and
$_{out}\!<K_L\ |\ K_S>_{in} \not = 0$, unless (\ref{eq:condTCP}) is
satisfied; this differs from
\cite{AlvarezGaumeKounnasLolaPavlopoulos} (see in particular (13) and
the end of section 4. We come back to CPLEAR in subsection
\ref{subsection:semilep}).}
.
One can check easily that ``in'' and ``out'' eigenstates match when the
propagator is also normal, that is when $|b|=|c|$ at $z=z_1$ and $z=z_2$:
one writes the kets for the ``out'' eigenstates, for example
$|\ K_L>\!_{out} = -\frac{i}{\sqrt{2\overline{a(z_1)}}}
\left( \sqrt{\overline{c(z_1)}}|\ K^0> - \sqrt{\overline{b(z_1)}}|\
\overline{K^0}> \right)$,
from which the results immediately follows.

\subsubsection{$\boldsymbol{CP}$ violating parameters}
\label{subsub:CPparam}

To get the indirect $CP$ violating parameters of all eigenstates in
(\ref{eq:ev4}) it is enough to go to the basis of $CP$ eigenstates
(\ref{eq:eigenCP3}). One defines the $\epsilon^{out}$ parameters {\em
for the kets} and not for the bras, which introduces complex conjugation of the
coefficients $b$ and $c$ (see subsection \ref{subsub:TCPeig} above). One gets

\vbox{
\begin{eqnarray}
&&\epsilon_L^{in} = \frac{\sqrt{b(z_1)} -e^{-i\alpha}\sqrt{c(z_1)}}
{\sqrt{b(z_1)} + e^{-i\alpha}\sqrt{c(z_1)}} ,\quad
\epsilon_S^{in} = \frac
{\sqrt{b(z_2)} - e^{-i\alpha}\sqrt{c(z_2)}}
{\sqrt{b(z_2)} +e^{-i\alpha}\sqrt{c(z_2)}}
,\cr
&&\cr
&&\epsilon_{\tilde L}^{in} = \frac{\sqrt{b(z_1)} +e^{-i\alpha}\sqrt{c(z_1)}}
{\sqrt{b(z_1)} - e^{-i\alpha}\sqrt{c(z_1)}},\quad
\epsilon_{\tilde S}^{in} = \frac
{\sqrt{b(z_2)} + e^{-i\alpha}\sqrt{c(z_2)}}
{\sqrt{b(z_2)} - e^{-i\alpha}\sqrt{c(z_2)}}
,\cr
&&\cr
&&\epsilon_L^{out} = \frac{\sqrt{\overline{c(z_1)}}
-e^{-i\alpha}\sqrt{\overline{b(z_1)}}}
{\sqrt{\overline{c(z_1)}} + e^{-i\alpha}\sqrt{\overline{b(z_1)}}},\quad
\epsilon_S^{out} = \frac
{\sqrt{\overline{c(z_2)}} - e^{-i\alpha}\sqrt{\overline{b(z_2)}}}
{\sqrt{\overline{c(z_2)}} +e^{-i\alpha}\sqrt{\overline{b(z_2)}}}
,\cr
&&\cr
&&\epsilon_{\tilde L}^{out} = \frac{\sqrt{\overline{c(z_1)}}
+e^{-i\alpha}\sqrt{\overline{b(z_1)}}}
{\sqrt{\overline{c(z_1)}} - e^{-i\alpha}\sqrt{\overline{b(z_1)}}},\quad
\epsilon_{\tilde S}^{out} = \frac
{\sqrt{\overline{c(z_2)}} + e^{-i\alpha}\sqrt{\overline{b(z_2)}}}
{\sqrt{\overline{c(z_2)}} -e^{-i\alpha}\sqrt{\overline{b(z_2)}}}
.
\label{eq:epsilons}
\end{eqnarray}
}
and one has the relations
\begin{eqnarray}
&&\epsilon_L^{in} = -\overline{\epsilon_L^{out}},\quad
\epsilon_S^{in} = -\overline{\epsilon_S^{out}}, \quad
\epsilon_{\tilde L}^{in} = -\overline{\epsilon_{\tilde L}^{out}},\quad
\epsilon_{\tilde S}^{in} = -\overline{\epsilon_{\tilde S}^{out}}, \cr
&& \epsilon_L^{in} = \frac{1}{\epsilon_{\tilde L}^{in}},\quad
\epsilon_S^{in} = \frac{1}{\epsilon_{\tilde S}^{in}},\quad
\epsilon_L^{out} = \frac{1}{\epsilon_{\tilde L}^{out}},\quad
\epsilon_S^{out} = \frac{1}{\epsilon_{\tilde S}^{out}}.
\label{eq:releps}
\end{eqnarray}
from the first line of which one gets in particular
\begin{equation}
|\epsilon_L^{in}|^2 = |\epsilon_L^{out}|^2,\
|\epsilon_S^{in}|^2 = |\epsilon_S^{out}|^2,\
|\epsilon_{\tilde L}^{in}|^2 = |\epsilon_{\tilde L}^{out}|^2,\
|\epsilon_{\tilde S}^{in}|^2 = |\epsilon_{\tilde S}^{out}|^2.
\label{eq:releps2}
\end{equation}
It is important to determine explicitly the real and imaginary parts of
the $\epsilon$'s, and to investigate how they change by a rephasing of the
$K^0$ and $\overline{K^0}$ fields
\begin{equation}
\varphi_{K^0} \rightarrow e^{i\omega}\varphi_{K^0},\quad
\varphi_{\overline{K^0}} \rightarrow e^{-i\omega}\varphi_{\overline{K^0}}.
\label{eq:rephase}
\end{equation}
We do this explicitly for $\epsilon_L^{in}$ and $\epsilon_L^{out}$.
Since the operator $\varphi_{K^0}$ annihilates the state $|\ K^0>$ to give
the vacuum, (\ref{eq:rephase}) entails that the states $|\ K^0>$
and $|\ \overline{K^0}>$ are re-phased by
\begin{equation}
|\ K^0> \rightarrow e^{-i\omega}|\ K^0>,\quad
|\ \overline{K^0}> \rightarrow e^{i\omega} |\ \overline{K^0}>.
\label{eq:rephase2}
\end{equation}
The way $\epsilon_L^{in}$ in (\ref{eq:epsilons}) is modified by $\omega$
is obtained by considering the first line of (\ref{eq:ev4}): it is
equivalent to changing in the expression (\ref{eq:epsilons})
 for $\epsilon_L^{in}$ $\sqrt{b(z_1)}$ into $e^{-i\omega}\sqrt{b(z_1)}$ and
$\sqrt{c(z_1)}$ into $e^{i\omega}\sqrt{c(z_1)}$;
for $\epsilon_L^{out}$, one finds that the same transformations are needed.
(\ref{eq:epsilons}) for $\epsilon_L^{in}$ and $\epsilon_L^{out}$ are
accordingly replaced by
\begin{equation}
\epsilon_L^{in} = \frac{\sqrt{b(z_1)} -e^{-i(\alpha-2\omega)}\sqrt{c(z_1)}}
{\sqrt{b(z_1)} + e^{-i(\alpha-2\omega)}\sqrt{c(z_1)}} ,\quad
\epsilon_L^{out} = \frac{\sqrt{\overline{c(z_1)}}
-e^{-i(\alpha-2\omega)}\sqrt{\overline{b(z_1)}}}
{\sqrt{\overline{c(z_1)}} + e^{-i(\alpha-2\omega)}\sqrt{\overline{b(z_1)}}}.
\label{eq:epsilons2}
\end{equation}
Writing
\begin{equation}
b(z_1) = |b_1|e^{i\beta_1},\quad c(z_1) = |c_1|e^{i\gamma_1},\quad
\Omega_1 = \alpha - 2\omega + \frac{\beta_1-\gamma_1}{2},
\label{eq:nota6}
\end{equation}
one obtains
\begin{equation}
\epsilon_L^{in} = \frac{|b_1| - |c_1| +2i\sqrt{|b_1c_1|} \sin\Omega_1}
{|b_1| + |c_1| +2\sqrt{|b_1c_1|} \cos\Omega_1},\quad
\epsilon_L^{out} = \frac{|c_1| - |b_1| +2i\sqrt{|b_1c_1|} \sin\Omega_1}
{|b_1| + |c_1| +2\sqrt{|b_1c_1|} \cos\Omega_1},
\label{eq:epsilons3}
\end{equation}
and, for their moduli
\begin{equation}
|\epsilon_L^{in}|^2 = |\epsilon_L^{out}|^2 =
 \frac{|b_1| + |c_1| -2\sqrt{|b_1c_1|} \cos\Omega_1}
{|b_1| + |c_1| +2\sqrt{|b_1c_1|} \cos\Omega_1},
\label{eq:modeps}
\end{equation}
which satisfy, when $\omega$ varies from $-\pi$ to $+\pi$
\begin{equation}
\left|\frac{\sqrt{|b_1|}-\sqrt{|c_1|}}{\sqrt{|b_1|}+\sqrt{|c_1|}}\right|\leq
|\epsilon_L^{in}| = |\epsilon_L^{out}| \leq
\left|\frac{\sqrt{|b_1|}+\sqrt{|c_1|}}{\sqrt{|b_1|}-\sqrt{|c_1|}}\right|.
\label{eq:boundmod}
\end{equation}
We observe that:\newline\null
- the real and imaginary parts of $\epsilon_L^{in}$ and $\epsilon_L^{out}$
and their moduli depend on the arbitrary phases $\omega$ and
$\alpha$;\newline\null
- the imaginary parts of $\epsilon_L^{in}$ and $\epsilon_L^{out}$ can
  always both be turned to $0$ by tuning $\omega$ (or
$\alpha$);\newline\null
- their real parts can never be cast to $0$ by such rephasing;\newline\null
- the real parts of $\epsilon_L^{in}$ and $\epsilon_L^{out}$ are opposite;
  their imaginary parts are identical;\newline\null
- the modulus $\left|\frac{1-\epsilon}{1+\epsilon}\right|$ is invariant by
  rephasing;\newline\null
- a variation of $\alpha$ can always compensate a variation of
  $\omega$;\newline\null
- when $|b|=|c|$, $\epsilon$ becomes purely imaginary, as already
  mentioned (see subsection \ref{subsub:normalpropag}).

The relations (\ref{eq:releps}) are unchanged by the rephasing
(see also appendix \ref{section:epsilon}).

When the arbitrary phases are varied, $\epsilon_L^{in}$ and
$\epsilon_L^{out}$ are located on two ellipsoids symmetric with respect to
the imaginary axis, as described in Fig.~2 of Appendix
\ref{section:epsilon}. The other $CP$ violating parameters are also
discussed there.

Accordingly, {\em a priori}, neither the real, nor the imaginary part, nor
the modulus of the $\epsilon$'s are physically relevant;
the only physical quantities are the lower and upper bounds
(\ref{eq:boundmod}) for the modulus of $\epsilon$; the upper bound being
much larger than $1$ can reasonably be discarded.
Nevertheless, as soon as $b_1 \not = c_1$, neither the real nor the modulus
of $\epsilon$ is vanishing; a non-zero  measurement of these
is accordingly a proof of $CP$ (or $T$) violation (see subsection
\ref{subsub:CPLEAR}).

The identification of the physically relevant quantity smooths out
the potential problems linked with the existence of two sets of physical
eigenstates, ``out'' and ``in'' which only differ by the signs of the real
parts of their $CP$ violating parameters
\footnote{
Section \ref{section:exper} will also show that no ambiguity arises in the
calculation of semi-leptonic asymmetries, which write in terms of
$\epsilon_S^{in}$ and $\epsilon_L^{in}$ only.}
.

Complements, in particular the comparison with the other $\epsilon$'s, can
be found is Appendix \ref{section:epsilon}.

\subsubsection{Expression of the eigenstates in terms of
the $\boldsymbol{CP}$ violating parameters}
\label{subsub:eigenstates}

In order to perform calculations of kaon decay amplitudes we will
need the expressions for the propagating states $K_L$ and $K_S$
in terms of the states with definite strangeness, $K_0$ and $\overline{K_0}$. Using
(\ref{eq:eigenCP3}, \ref{eq:ev4}, \ref{eq:normeig},
\ref{eq:releps}) we obtain
\footnote{It is instructive to compare (\ref{eq:CPeigeps}) with (15) and (16)
in \cite{AlvarezGaumeKounnasLolaPavlopoulos}.}
\begin{eqnarray}
|K_L>_{in} & = & \frac{1}{\sqrt 2}
\left[\sqrt{\frac{1+\epsilon_L^{in}}{1-\epsilon_L^{in}}} |\ K_0> -
e^{i\alpha}\sqrt{\frac{1-\epsilon_L^{in}}{1+\epsilon_L^{in}}} |\ \overline{K_0}>
\right]
=\frac{1}{\sqrt 2}\left[\frac{1}{\xi_L}|\ K^0> 
-e^{i\alpha}\xi_L|\ \overline{K^0}> \right],\cr
 |K_S>_{in} & = & \frac{1}{\sqrt 2}
\left[\sqrt{\frac{1+\epsilon_S^{in}}{1-\epsilon_S^{in}}} |\ K_0> +
e^{i\alpha}\sqrt{\frac{1-\epsilon_S^{in}}{1+\epsilon_S^{in}}} |\ \overline{K_0}>
\right]
=\frac{1}{\sqrt 2}\left[\frac{1}{\xi_S}|\ K^0>+e^{i\alpha}\xi_S
 |\ \overline{K^0}> \right],\cr
_{out}\!<K_L| & = & \frac{1}{\sqrt 2}
\left[\sqrt{\frac{1-\epsilon_L^{in}}{1+\epsilon_L^{in}}} <K_0\ | -
e^{-i\alpha}\sqrt{\frac{1+\epsilon_L^{in}}{1-\epsilon_L^{in}}} <\overline{K_0}\ |
\right]
=\frac{1}{\sqrt 2}\left[\xi_L <K^0\ |- e^{-i\alpha}\frac{1}{\xi_L}
<\overline{K^0}\
| \right],\cr
_{out}\!<K_S| & = & \frac{1}{\sqrt 2}
\left[\sqrt{\frac{1-\epsilon_S^{in}}{1+\epsilon_S^{in}}} <K_0\ | +
e^{-i\alpha}\sqrt{\frac{1+\epsilon_S^{in}}{1-\epsilon_S^{in}}} <\overline{K_0}\ |
\right]
=\frac{1}{\sqrt 2}\left[\xi_S<K^0\ |+e^{-i\alpha} \frac{1}{\xi_S} <\overline{K^0}\
|\right],\cr
&&
\label{eq:CPeigeps}
\end{eqnarray}
where we have introduced the notations
\begin{equation}
\xi_S = \sqrt{\frac{1-\epsilon_S^{in}}{1+\epsilon_S^{in}}} \equiv
\sqrt{\sqrt{\frac{c(z_2)}{b(z_2)}}e^{-i\alpha}},\quad
\xi_L = \sqrt{\frac{1-\epsilon_L^{in}}{1+\epsilon_L^{in}}} \equiv
\sqrt{\sqrt{\frac{c(z_1)}{b(z_1)}}e^{-i\alpha}}.
\label{eq:nota5}
\end{equation}
$|\xi_L|$, $|\xi_S|$ and $\frac{\xi_L}{\xi_S}$ are invariant by
(\ref{eq:rephase2}).

Inverting (\ref{eq:CPeigeps}) one obtains
\footnote{ $|\ K_L>_{in}$, $|\ K_L>_{out}$, $|\ K_S>_{in}$ and $|\ K_S>_{out}$
are not linearly independent; since \cite{Machet}, unlike in
\cite{AlvarezGaumeKounnasLolaPavlopoulos}, there is no distinction
between $|\ K^0>_{in}$ and $|\ K^0>_{out}$,
(\ref{eq:flaveigeps}) determine
$|\ K_S>_{out}$ and $|\ K_L>_{out}$ as linear combinations
of $|\ K_S>_{in}$ and $|\ K_L>_{in}$.
}
:

\vbox{
\begin{eqnarray}
|\ K^0> &=& \frac{1}{D\sqrt{2}}\left[
\sqrt{\frac{1-\epsilon_S^{in}}{1+\epsilon_S^{in}}}|\ K_L>_{in}
+\sqrt{\frac{1-\epsilon_L^{in}}{1+\epsilon_L^{in}}}|\ K_S>_{in}
\right]
=\frac{1}{D\sqrt{2}}\Big[\xi_S |\ K_L>_{in} + \xi_L|\ K_S>_{in}
\Big],\cr
|\ \overline{K^0}> &=& \frac{e^{-i\alpha}}{D\sqrt{2}}\left[
-\sqrt{\frac{1+\epsilon_S^{in}}{1-\epsilon_S^{in}}}|\ K_L>_{in}
+\sqrt{\frac{1+\epsilon_L^{in}}{1-\epsilon_L^{in}}}|\ K_S>_{in}
\right]
=\frac{e^{-i\alpha}}{D\sqrt{2}}\left[-\frac{1}{\xi_S}|\ K_L>_{in}
+ \frac{1}{\xi_L}|\ K_S>_{in}
\right],\cr
<K^0\ | &=& \frac{1}{D\sqrt{2}}\left[
\sqrt{\frac{1+\epsilon_S^{in}}{1-\epsilon_S^{in}}}\;{ _{out}\!<K_L\ |}
+\sqrt{\frac{1+\epsilon_L^{in}}{1-\epsilon_L^{in}}}\;{ _{out}\!<K_S\ |}
\right]
=\frac{1}{D\sqrt{2}}\left[\frac{1}{\xi_S}\; {_{out}\!<K_L\ |} +
\frac{1}{\xi_L}
 \;{_{out}\!<K_S\ |}
\right],\cr
<\overline{K^0}\ | &=& \frac{e^{i\alpha}}{D\sqrt{2}}\left[
-\sqrt{\frac{1-\epsilon_S^{in}}{1+\epsilon_S^{in}}}\; {_{out}\!<K_L\ |}
+\sqrt{\frac{1-\epsilon_L^{in}}{1+\epsilon_L^{in}}}\; {_{out}\!<K_S\ |}
\right]
=\frac{e^{i\alpha}}{D\sqrt{2}}\Big[-\xi_S\; {_{out}\!<K_L\ |} + \xi_L\; {_{out}\!<K_S\
|} \Big],\cr
&&
\label{eq:flaveigeps}
\end{eqnarray}
}
where 
\begin{equation}
D =
\frac{1}{2}\left(
\sqrt{\frac{(1+\epsilon_L^{in})(1-\epsilon_S^{in})}{(1-\epsilon_L^{in})(1+\epsilon_S^{in})}}
+\sqrt{\frac{(1+\epsilon_S^{in})(1-\epsilon_L^{in})}{(1-\epsilon_S^{in})(1+\epsilon_L^{in})}}
\right)=
\frac{1-\epsilon_L^{in}\epsilon_S^{in}}{\sqrt{(1-(\epsilon_L^{in})^2)(1-(\epsilon_S^{in})^2)}}.
\label{eq:D}
\end{equation}

\subsubsection{$\boldsymbol{TCP}$ invariance is realized in a non-trivial way}
\label{subsub:realTCP}

From  (\ref{eq:ev4}) and the relations (\ref{eq:releps}), one concludes
that
$TCP$ symmetry is realized at each given $z$, among the two corresponding
``in'' eigenstates (one physical and one ``spurious''),
 and, likewise, among the two ``out'' eigenstates;
$|\ K_L>_{in}$ and $|\ \tilde K_L>_{in}$ have $CP$-violating parameters
satisfying $\epsilon_L^{in} = 1/\epsilon_{\tilde L}^{in}$
\footnote{The fact that these two $CP$-violating parameters are inverse of
each other instead of being identical is only due to the convention that we
have chosen, and it is indeed a consequence
of $TCP$ invariance.}
; the same type of relations occurs between $|\ K_S>_{in}$ and $|\ \tilde
K_S>_{in}$, and between the two similar pairs of ``out'' states.

 This means in
particular that, at any of the two physical masses, $TCP$ symmetry needs
both the propagating and spurious eigenstates to be realized.

However this does not occur in general for the physical propagating $K_L$ and
$K_S$ mesons, because they correspond to two different values of $z$
\begin{equation}
\text{For}\ z_1 \not=z_2,\quad \epsilon_L^{in} \not=
{\epsilon_S^{in}},\quad
\epsilon_L^{out} \not= {\epsilon_S^{out}};
\label{eq:diffeps}
\end{equation}
the equalities are satisfied only when (see (\ref{eq:epsilons}))
\begin{equation}
\frac{b(z_1)}{c(z_1)} = \frac{b(z_2)}{c(z_2)}=\frac{1}{\zeta}.
\label{eq:condTCP}
\end{equation}
(\ref{eq:condTCP}) transcribed for the elements of the propagator,
instead of the inverse propagator writes
\begin{equation}
\frac{g(z_1)}{h(z_1)} = \frac{g(z_2)}{h(z_2)};
\label{eq:condTCP2}
\end{equation}
a particular cases when it is satisfied are when the two physical masses
are identical $z_1=z_2$ (a trivial one if of course when $CP$ invariance
holds, as (\ref{eq:CPcons1}) tells us, since the phase $\alpha$ is a
constant);

(\ref{eq:condTCP})(\ref{eq:condTCP2}) are, in general,
not fulfilled, such that
the physical mass eigenstates do not satisfy the most commonly used
criterion of  $TCP$ invariance;
of course the $TCP$ symmetry is achieved and stays a fundamental
property of the theory.

We shall investigate in sections \ref{section:massmatrix} and
\ref{section:exper} what are
the consequences on the mass matrix, and if effects which would mimic
 $TCP$ violating can be expected in experiments and wrongly interpreted as
a violation of this fundamental symmetry.

\subsubsection{Bi-unitary transformations}
\label{subsub:nobiunit}

We should discriminate between the two ways of diagonalizing the
propagator: using a biunitary transformation or a bi-orthogonal basis
 (see subsection \ref{subsub:diago}).

We will compare below the two procedures at the poles;
this will give us a criterion to reject bi-unitary
transformations.

For this, we shall suppose that $\Delta^{-1}(z)$ is not normal at the poles,
 too, and we shall explicitly calculate the eigenvectors obtained by a
bi-unitary transformation.

If $\det(\Delta^{-1}(z)=0$, the determinant of
$\frac{1}{\Delta(z)}\frac{1}{(\Delta(z))^\dagger}$ vanishes, too,
and so does the
determinant of   $\frac{1}{(\Delta(z))^\dagger}\frac{1}{\Delta(z)}$; so,
these three sets of functions have poles at the same locations: the
physical masses are the same in the two procedures.

At these poles $z=z_1$ and $z=z_2$,
$\Delta^{-1}(z)$ can be written, using (\ref{eq:physmass3})
\begin{equation}
\Delta^{-1}(z) = \left( \begin{array}{cc}
                 \pm \sqrt{b(z)c(z)}  &  -b(z) \cr
                      -c(z)         & \pm \sqrt{b(z) c(z)}
\end{array}\right)\ \text{at}\ z=z_1\ \text{or}\ z=z_2.
\label{eq:pp1}
\end{equation}
From (\ref{eq:pp1}) one gets immediately $(\Delta^{-1}(z))^\dagger$, and the
roots of the characteristic equation of
$\Delta^{-1}(z) (\Delta^{-1}(z))^\dagger$ or $(\Delta^{-1}(z))^\dagger
\Delta^{-1}(z)$ are  found to be
\begin{equation}
\delta = 0,\  \zeta = (|b(z)| + |c(z)|)^2,
\end{equation}
where $z$ is to be considered to be equal to $z_1$ or to $z_2$.

The two unitary matrices $U$ and $V$ which are used to diagonalize
$\Delta^{-1}$, and which respectively diagonalize the hermitian matrices
$\Delta^{-1}(z) [\Delta^{-1}(z)]^\dagger$ and $[\Delta^{-1}(z)]^\dagger
\Delta^{-1}(z)$ are accordingly constructed from the eigenvectors of
$\Delta^{-1}(z) [\Delta^{-1}(z)]^\dagger$ and $[\Delta^{-1}(z)]^\dagger
\Delta^{-1}(z)$, which form two different orthonormal basis.

The eigenvectors we characterize as usual by the ratio $v/u$
 of their components in the $(K^0, \overline{K^0})$ basis.

$\bullet$\ The eigenvectors of $\Delta^{-1}(z) [\Delta^{-1}(z)]^\dagger$ are the
following:\newline\null
\hskip 2cm$\ast$ at $z=z_1$;\newline\null
\hskip 3.5cm - for the vanishing eigenvalue $\frac{v}{u}(z_1) =
   - \sqrt{\frac{\overline{ b(z_1)}}{\overline{ c(z_1)}}}$;\newline\null
\hskip 3.5cm  - for the eigenvalue $(|b(z_1)| + |c(z_1)|)^2$,
$\frac{v}{u}(z_1) = +\sqrt{\frac{c(z_1)}{b(z_1)}}$;\newline\null
\hskip 2cm$\ast$ at $z=z_2$;\newline\null
\hskip 3.5cm - for the vanishing eigenvalue $\frac{v}{u}(z_2) =
   + \sqrt{\frac{\overline{ b(z_2)}}{\overline{ c(z_2)}}}$;\newline\null
\hskip 3.5cm  - for the eigenvalue $(|b(z_2)| + |c(z_2)|)^2$,
$\frac{v}{u}(z_2) = -\sqrt{\frac{c(z_2)}{b(z_2)}}$;

$\bullet$\ the eigenvectors of $[\Delta^{-1}(z)]^\dagger (\Delta^{-1}(z))$ are the
following:\newline\null
\hskip 2cm$\ast$ at $z=z_1$;\newline\null
\hskip 3.5cm - for the vanishing eigenvalue $\frac{v}{u}(z_1) =
   - \sqrt{\frac{c(z_1)}{b(z_1)}}$;\newline\null
\hskip 3.5cm  - for the eigenvalue $(|b(z_1)| + |c(z_1)|)^2$,
$\frac{v}{u}(z_1) = +\sqrt{\frac{\overline{ b(z_1)}}{\overline{
c(z_1)}}}$;\newline\null
\hskip 2cm$\ast$ at $z=z_2$;\newline\null
\hskip 3.5cm - for the vanishing eigenvalue $\frac{v}{u}(z_2) =
   + \sqrt{\frac{c(z_2)}{b(z_2)}}$;\newline\null
\hskip 3.5cm  - for the eigenvalue $(|b(z_2)| + |c(z_2)|)^2$,
$\frac{v}{u}(z_2) = -\sqrt{\frac{\overline{ b(z_2)}}{\overline{ c(z_1)}}}$.

Comparing the {formul\ae}\,  above with (\ref{eq:ev4}), we
conclude that:\newline\null
$\ast$ the  eigenvectors of $(\Delta^{-1}(z))^\dagger \Delta^{-1}(z)$
for the vanishing eigenvalues match
the ``in'' propagating states of (\ref{eq:ev4});\newline\null
$\ast$ the  eigenvectors of $\Delta^{-1}(z)(\Delta^{-1}(z))^\dagger$
for the vanishing eigenvalues match the {\em kets}
\footnote{Going from the ``bras'' of (\ref{eq:ev4}) to the corresponding
``kets'' yields the complex conjugation of
coefficients, which provides the matching.}
 corresponding to
the ``out'' propagating states of (\ref{eq:ev4});\newline\null
$\ast$ the  eigenvectors of $(\Delta^{-1}(z))^\dagger \Delta^{-1}(z)$
for the non-vanishing eigenvalues only match the ``in''
 spurious states of (\ref{eq:ev4}) when the propagator is normal
($|b|=|c|$);\newline\null
$\ast$ the  eigenvectors of $\Delta^{-1}(z) (\Delta^{-1}(z))^\dagger$
for the non-vanishing eigenvalues only match the ``kets'' corresponding to
the ``out'' spurious states of (\ref{eq:ev4}) when the propagator
is normal ($|b|=|c|$).

We have thus shown that the difference between a bi-unitary diagonalization
and the use of a bi-orthogonal basis lies, at $z$ equal to the physical
masses, in the spurious states; while the propagating states match
correctly, the other ones differ, unless the propagator is normal; this is
as expected since a biunitary transformation provides an orthogonal basis of
eigenvectors in each space, ``in'' and ``out'', while the other procedure
only provides a bi-orthogonal basis spanning the two spaces
\footnote{We thus disagree with footnote 1 of
\cite{AlvarezGaumeKounnasLolaPavlopoulos}; however, the matrices which
connect the eigenstates of the Hamiltonian with the $K^0$ and $\overline{K^0}$
fields according to the bulk of the paper
\cite{AlvarezGaumeKounnasLolaPavlopoulos} are not unitary.}
.

\subsection{UNITARITY}
\label{subsection:unitarity}

The argument that follows exists independently of the instability of the
physical eigenstates and is not related to the non-hermiticity of the
associated Lagrangian. It only relies on the existence of a mass splitting.

We shall accordingly first consider the case of a normal Lagrangian,
which can be diagonalized by a single unitary transformation; ``in'' and
``out'' eigenstates coincide, and form, at any value of $q^2$
independent orthonormal basis.
The $q^2$ dependent matrix that relates flavor to $q^2$
dependent eigenstates is
unitary, and the unitarity of the theory cannot be cast in doubt.

However, as soon as a binary system is mass split,
its propagator evaluated at the
two different physical masses differ, and the occurrence of spurious states
is unavoidable, which are essential to complete, at any of the two physical
poles $z_1$ and $z_2$, the corresponding orthonormal basis.
If one evaluate flavor states $K^0$ and $\overline{K^0}$ in terms of the two
physical'' propagating states, the corresponding ``mixing'' matrix can no
longer be unitary, because its columns are evaluated respectively at
$z=z_1$ and $z=z_2$.

A deeper investigation of this phenomenon and its consequences for the
mixing matrix will be performed in \cite{MachetNovikovVysotsky}.

\subsection{THE SPECIAL CASE $\boldsymbol{\epsilon_L = \epsilon_S}$}
\label{subsection:epsequal}

(\ref{eq:condTCP}) and the equality between $\epsilon_L$ and $\epsilon_S$
are not forced by the $TCP$ symmetry; in the quark model,
an estimate of their difference is presented
in subsection \ref{subsub:deltaeps}.

It is nevertheless instructive to describe the simplifications that occur
when their equality is assumed (this is the usual situation in QM).

(\ref{eq:condTCP}) applied to (\ref{eq:ev4}) shows
that the spurious eigenstates
``disappear'' by becoming identical to the mass eigenstates $K_L$ and $K_S$;
the picture that arises is the following:

- at $z=z_1$, the eigenstates of the propagator (and of its inverse)
are $K_L$ and $K_S$; $K_L$ propagates (at $z=z_1$ the inverse
propagator vanishes) but $K_S$ does not, since it does not correspond to a
pole of the propagator;

- at $z=z_2$ the reverse occurs: the eigenstates of $\Delta$ and
  $\Delta^{-1}$ are again $K_L$ and $K_S$ but, now, $K_S$ propagates and
$K_L$ does not.

The set of $K_L$ and $K_S$ eigenstates satisfy 
${_{out}<K_L(t)\ |\ K_S(t)>_{in}} = 0 = {_{out}<K_S(t)\ |\ K_L(t)>_{in}}$
for all $t$, which is not true in the general case.

As shown in subsection \ref{subsub:charas} measuring
the difference between the semi-leptonic asymmetries $\delta_L - \delta_S$
amounts to  a test of the non-vanishing of $\epsilon_L -\epsilon_S$.

\section{PROPERTIES AND LIMITATIONS OF MASS MATRICES}
\label{section:massmatrix}

When, in textbooks of Quantum Mechanics or Quantum Field Theory
a mass matrix is introduced for neutral kaons, it is a constant matrix
\footnote{We limit our discussion to constant mass matrices. Results similar to
ours (the non-equality of $CP$ violating parameters for $K_L$ and $K_S$
despite $TCP$ is satisfied) had previously been obtained in \cite{Azimov},
in the formalism of an energy-dependent Hamiltonian.
}
(in the bare Lagrangian)
\footnote{Its renormalization is still subject to many
debates (see for example \cite{Pilaftsis} and references therein).}
.
We have seen in subsection \ref{subsub:massmat} that introducing a constant
mass matrix can only  be an
approximation, valid when a linear expansion of the inverse
propagator is suitable (most likely very close to the poles).

The link should nevertheless be made with such a matrix since, in particular,
all experiments are analyzed and fitted with the corresponding parameters.

The question of its normality is rapidly settled: since the propagator
cannot be normal, in particular on the cut, the mass matrix cannot either.
Also, from our general discussion on normal matrices in subsection
\ref{subsub:normalpropag}, it is clear that the mass matrix ${M^{\{2\}}}$ (see
(\ref{eq:mama})) cannot
be normal since this would lead to purely imaginary indirect $CP$
violating parameters.

Non-normal mass matrices have different left and right eigenstates; the
corresponding question of knowing which kind of
eigenstate is detected, was up to now left unsolved, often qualified of
``unavoidable mathematical necessity''.
We showed in subsection \ref{subsub:CPparam} that the $CP$ violating
parameters of the ``in'' and ``out'' states corresponding to the same pole
of the propagator have equal imaginary parts and opposite real parts, but
this distinction does not appear physically relevant since both quantities turn
out to depend on an arbitrary rephasing of $K^0$ (and $\overline{K^0}$). This
question is finally wiped out in section \ref{section:exper} where we show
that semi-leptonic asymmetries can be unambiguously calculated in terms of
the sole $\epsilon_L^{in}$ and $\epsilon_S^{in}$.

The determination of the mass matrix is the question that we address now.
Can one define a unique constant  mass
matrix which has the correct eigenmasses and for eigenstates the correct
propagating eigenstates that we have rigorously defined above?

\subsection{THE MASS MATRIX IN QUANTUM MECHANICS}
\label{subsection:QM}

In QM, one introduces the complex mass matrix $\cal M$
\footnote{See also footnote \ref{footnote:Silva}.}
 with
dimension $[mass]$
\begin{eqnarray}
{\cal M} &=&
M-\frac{i}{2} \Gamma,\ \text{with}\ M=M^\dagger,\ \Gamma=\Gamma^\dagger,\cr
&=& \left( \begin{array}{cc}
   m_{11} -\frac{i}{2} \gamma_{11}  &  m_{12} -\frac{i}{2} \gamma_{12}\cr
  \overline{m_{12}} -\frac{i}{2}\overline{\gamma_{12}} & m_{22}-\frac{i}{2}\gamma_{22}
\end{array}\right),\ \text{with}\ m_{11}, m_{22},\gamma_{11},\gamma_{22}
\in{\mathbb R}.
\label{eq:MMQ}
\end{eqnarray}
$\cal M$ is normal if and only if $M$ and $\Gamma$ commute,
$[M,\Gamma]=0$.

The conditions of $TCP$ invariance are $M_{11}=M_{22}$
\cite{Okun1}\cite{BrancoLavouraSilva}, and  $CP$
invariance adds to it the condition
$M_{12} = e^{-2i\alpha} M_{21}$, such that a $CP$ invariant mass matrix
is always of the form
\begin{equation}
{\cal M}_{CP} =
\left( \begin{array}{cc}
        m -\frac{i}{2} \gamma  &  e^{-i\alpha}(\omega -\frac{i}{2} \chi)\cr
        e^{i\alpha}(\omega -\frac{i}{2}\chi) & m-\frac{i}{2}\gamma
\end{array}\right),\ \text{with}\ m,\gamma,\omega,\chi \in {\mathbb R},
\label{eq:MMQCP}
\end{equation}
which is a normal matrix.

Since the mass matrix is supposed to describe unstable kaons, it cannot be
hermitian, because its eigenvalues, which are the masses of
the eigenstates, would then be real.

Experiments tell us that the mass matrix of neutral $K$ mesons
 cannot even be normal
\footnote{$K^0 \rightarrow \overline{K^0}$ probability is different from
$\overline{K^0} \rightarrow K^0$ probability and
the corresponding $CP$ violating parameter
cannot be purely imaginary; this entails that the mass matrix cannot be
normal (see also subsection \ref{subsub:normalpropag}).}
 and thus
should be diagonalized either by a bi-unitary transformation, or by using a
bi-orthogonal basis. Since bi-unitary transformations always yield real
masses, they are excluded for the same reasons as mentioned above.
As done in \cite{AlvarezGaumeKounnasLolaPavlopoulos}, one must use a
bi-orthogonal basis
\footnote{The criterion of masses real or not is only valid for a mass
matrix and not for the full propagator. Indeed, when $\lambda_\pm(z)$ are
the eigenvalues of the propagator, their reality does not prevent the
physical masses, which are the solutions of $\lambda_\pm(z)=0$ to be
complex. Hence, we rejected bi-unitary transformations for the
propagator with more involved arguments.}
.

\subsection{INCONSISTENCY OF A CONSTANT MASS MATRIX IN QFT}
\label{subsection:diago}

Let ${M^{\{2\}}}$ be a constant complex matrix
\begin{equation}
{M^{\{2\}}} = \left( \begin{array}{cc}   n  & r \cr
                                s  & t \end{array}\right), n,r,s,t \in
{\mathbb C}.
\label{eq:cmm}
\end{equation}
We request that the exact propagating eigenstates $|\ K_L>_{in}$, $|\
K_S>_{in}$, $_{out}\!<K_L\ |$ and $_{out}\!<K_S\ |$ determined in
(\ref{eq:ev4}) be its eigenstates, and we forget about the
spurious states at $z=z_1$ and $z=z_2$ which cannot be accounted for
in this restricted formalism.

For the sake of simplicity we shall adopt the following notations
\footnote{Do not confuse the present notations $b_1$ and $c_1$ with the
ones used in (\ref{eq:nota6}).}
:
\begin{equation}
b_1 = \sqrt{\frac{b(z_1)}{2a(z_1)}},\
b_2 = \sqrt{\frac{b(z_2)}{2a(z_2)}},\
c_1= \sqrt{\frac{c(z_1)}{2a(z_1)}},\
c_2 = \sqrt{\frac{c(z_2)}{2a(z_2)}}.
\label{eq:not2}
\end{equation}
The eigenvalues of ${M^{\{2\}}}$ are
\begin{equation}
\mu_\pm = \frac{1}{2}\left( n+t \pm \sqrt{(n-t)^2 + 4rs}\right);
\label{eq:mupm}
\end{equation}
the equations for the eigenstates and their identification with the true
propagating states (\ref{eq:ev4}) write
\begin{eqnarray}
&&{M^{\{2\}}} \left(\begin{array}{c} u_+ \cr v_+\end{array}\right) =
\mu_+ \left(\begin{array}{c} u_+ \cr v_+\end{array}\right) =
i\mu_+ \left(\begin{array}{r} b_1 \cr
-c_1\end{array}\right), \cr
&&{M^{\{2\}}} \left(\begin{array}{c} u_- \cr v_-\end{array}\right) =
\mu_- \left(\begin{array}{c} u_- \cr v_-\end{array}\right) =
\mu_- \left(\begin{array}{r} b_2 \cr
c_2\end{array}\right), \cr
&&\left( x_+,\ y_+\right) {M^{\{2\}}} = \left( x_+, \ y_+\right) \mu_+
= \left(c_1,\ -b_1\right)i\mu_+,\cr
&&\left( x_-, \ y_-\right) {M^{\{2\}}} = \left( x_-, \ y_-\right) \mu_-
= \left(c_2,\ b_2\right) \mu_-,
\label{eq:ident}
\end{eqnarray}
which, using (\ref{eq:mupm}), leads to

\vbox{
\begin{eqnarray}
&&\frac{v_+}{u_+} = \frac{t-n + \sqrt{(t-n)^2 + 4rs}}{2r}
= \frac{2s}{n-t+ \sqrt{(t-n)^2 + 4rs}} = -\frac{c_1}{b_1} ,\quad(a)\cr
&&\frac{v_-}{u_-} = \frac{t-n - \sqrt{(t-n)^2 + 4rs}}{2r}
= \frac{2s}{n-t- \sqrt{(t-n)^2 + 4rs}} = \frac{c_2}{b_2} ,\quad(b)\cr
&&\frac{y_+}{x_+} = \frac{t-n + \sqrt{(t-n)^2 + 4rs}}{2s}
=\frac{2r}{n-t+ \sqrt{(t-n)^2 + 4rs}} = -\frac{b_1}{c_1},\quad(c)\cr
&&\frac{y_-}{x_-} = \frac{t-n - \sqrt{(t-n)^2 + 4rs}}{2s}
=\frac{2r}{n-t- \sqrt{(t-n)^2 + 4rs}} = \frac{b_2}{c_2}.\quad(d)
\label{eq:rats}
\end{eqnarray}
}
Equations (\ref{eq:rats}) are contradictory unless (\ref{eq:condTCP}) is
satisfied; indeed:\newline\null
- for (a) and (c) to be inverse of each other, as their r.h.s. demand, one
  needs the equality of diagonal elements $n=t$; the same occurs for
equations (b) and (d);\newline\null
- they also show that the two diagonal elements $n$ and $t$
 of ${M^{\{2\}}}$ cannot be identical unless (\ref{eq:condTCP}) is satisfied:
indeed, if they were, one would get
from (a) and (b) $\frac{v_+}{u_+}=-\frac{v_-}{u_-} =\sqrt{\frac{s}{r}}$,
which, from their r.h.s. entails $\frac{c_1}{b_1}=\frac{c_2}{b_2}$, which
is condition (\ref{eq:condTCP}); the same occurs between (c) and (d).

We conclude that one can introduce a constant mass matrix only with
constant diagonal matrix elements, which then can only correspond to the
case when $TCP$ symmetry is achieved between the propagating eigenstates
(the $CP$ violating parameters for $K_L$ and $K_S$ are then identical).

\subsection{CONSTANT MASS MATRIX AND DISCRETE SYMMETRIES}
\label{subsection:param}

Since (\ref{eq:condTCP}) has no reasons to be satisfied in general,
we reach the following conclusion:

{\em A constant mass matrix can never describe faithfully  the correct
propagating eigenstates of a quasi-degenerate system of neutral mesons};

indeed:\newline\null
-  choosing its diagonal elements equal is equivalent to imposing
$\epsilon_L =1/ \epsilon_S$, the equality of the $CP$ violating parameters of
$K_L$ and $K_L$, which we have seen to be untrue; setting different
diagonal elements leads to contradictions between (a) and (c), and (b) and
(d), which means that either the ``in'', or the ``out'' eigenstates of the
constant mass matrix match those of the propagator, but never
both;\newline\null
- it cannot include the spurious states, which means in particular
that any completeness relation obtained from ${M^{\{2\}}}$ is {\em a priori}
incorrect (it does not build the appropriate Hilbert spaces of
``in''; and ``out states).

A constant mass matrix is in particular inappropriate
to provide a faithful description of $TCP$ symmetry.

Can it describe faithfully $CP$ violation?

Giving up a correct description of  $TCP$, let us choose
${M^{\{2\}}}$ with equal diagonal elements $t=n$, which is the ``quantum mechanical
condition'' for $TCP$ invariance. To be consistent, one must
identify by brute force the indirect $CP$ violating
parameters of $K_L$ and $K_S$, which can only be done by assuming
(see subsection \ref{subsub:realTCP}) $\frac{b_1}{c_1} = \frac{b_2}{c_2}$;
(\ref{eq:rats}) entails furthermore that this is automatically
satisfied when $n=t$.
The number of parameters of ${M^{\{2\}}}$ to be determined shrinks to three; to fix
them we have at our disposal three equations, respectively for the two masses
and  for the (now unique) $CP$ violating parameter $\epsilon$ (the one for
``out'' eigenstates is easily deduced from the one for ``in'' eigenstates,
see (\ref{eq:epsilons}) (\ref{eq:releps}) (\ref{eq:releps2})).

We conclude that a constant mass
matrix has enough parameters to provide a faithful parameterization of $CP$
violation, or $T$ violation, once $TCP$ conservation is assumed and
imposed. The numerical accuracy of the constant mass matrix approximation
will be determined in subsection \ref{subsub:deltaeps}.

\section{PHYSICAL PROCESSES AND ASYMMETRIES}
\label{section:exper}

The goal of this section is to complete the previous formal
investigations by phenomenological considerations. In particular, we want to
stress  differences of interpretation between a description of
neutral mesons by QM and by QFT.

Since the main difference that we outlined above concern discrete
symmetries, in particular $TCP$, and the
difference of the $CP$ violating parameters $\epsilon_L \not = \epsilon_S$,
this will be our principal topic below.

It is important to notice that all asymmetries are expressed
in terms of the $CP$ violating parameters $\epsilon_S^{in}$ and
$\epsilon_L^{in}$; the parameters of the ``out'' states do not appear.
The distinction between ``in'' and ``out'' states has no consequences for
physical observables.

\subsection{SEMI-LEPTONIC ASYMMETRIES}
\label{subsection:semilep}

We shall calculate the following asymmetries \cite{RPP}
\begin{eqnarray}
A_T &=& \frac{|<\pi^- \ell^+ \nu (t_f)\ |\
\overline{K^0}(t_i)>|^2 - |<\pi^+ \ell^- \overline{\nu} (t_f)\ |\
{K^0}(t_i)>|^2} {|<\pi^- \ell^+ \nu (t_f)\ |\
\overline{K^0}(t_i)>|^2 + |<\pi^+ \ell^- \overline{\nu} (t_f)\ |\
{K^0}(t_i)>|^2}\cr
&\stackrel{\Delta S = \Delta Q}{=}&\frac{|<K^0(t_f)\ |\ \overline{K^0}(t_i)>|^2
-|<\overline{K^0}(t_f)\ |\ {K^0}(t_i)>|^2}
{|<K^0(t_f)\ |\ \overline{K^0}(t_i)>|^2
+|<\overline{K^0}(t_f)\ |\ {K^0}(t_i)>|^2};\cr
&&\cr &&\cr
\delta_{L,S}
&=& \frac{|_{out}\!<\pi^- \ell^+ \nu\ |\ K_{L,S}>_{in}|^2 -
|_{out}\!<\pi^+ \ell^- \overline{\nu}\ |\ K_{L,S}>_{in}|^2}
{|_{out}\!<\pi^- \ell^+ \nu\ |\ K_{L,S}>_{in}|^2 +
|_{out}\!<\pi^+ \ell^- \overline{\nu}\ |\ K_{L,S}>_{in}|^2};\cr
&&\cr &&\cr
A_{TCP} & = & \frac{|<\pi^+ \ell^- \bar\nu(t_f)| \bar
K^0(t_i)>|^2 - |<\pi^- \ell^+ \nu(t_f)| K^0>|^2}{|<\pi^+ \ell^-
\bar\nu(t_f)| \overline{K^0}(t_i)>|^2 + |<\pi^- \ell^+ \nu(t_f)| K^0>|^2}\cr
&&\cr
&\stackrel{\Delta S = \Delta Q}{=}&\frac{|<\overline{K^0}(t_f)\ |\ \overline{K^0}(t_i)>|^2
- |<{K^0}(t_f)\ |\ {K^0}(t_i)>|^2}
{|<\overline{K^0}(t_f)\ |\ \overline{K^0}(t_i)>|^2
+ |<{K^0}(t_f)\ |\ {K^0}(t_i)>|^2}.
\label{eq:asyms1}
\end{eqnarray}
supposing that the rule $\Delta S = \Delta Q$ holds
\footnote{It is expected from the standard model to be valid up to
order $10^{-14}$ \cite{AlvarezGaumeKounnasLolaPavlopoulos}.
\label{footnote:dSdQ}}
, which in particular
only allows the semi-leptonic decays
\begin{equation}
K^0 \rightarrow \pi^- \ell^+ \nu, \quad \overline{K^0} \rightarrow \pi^+ \ell^-
\overline{\nu}.
\label{eq:dSdQ}
\end{equation}
The first and the third are the asymmetries tested in the CPLEAR
experiment \cite{CPLEAR}, $\delta_L$ has been accurately measured
\cite{RPP} and $\delta_S$ should be measured with tagged $K_S$ at $\phi$ factories.

\subsubsection{The CPLEAR asymmetry $\boldsymbol{A_T}$ \cite{Pavlopoulos}}
\label{subsub:CPLEAR}

At low energy $\bar p p$ annihilation $K^- K^0$ ($K^+ \overline{K^0}$)
are produced; $K^0$ ($\overline{K^0}$) is tagged by $K^-$ ($K^+$) decay.
The particle momenta are not measured with high accuracy (in the opposite
case it would be known that $K_L$ -- or $K_S$ --  was produced).
As soon as an ``averaging'' over $K^0$ ($\overline{K^0}$) momentum is accepted, the
contributions of both $K_L$ and $K_S$ in intermediate (propagating) states
should be taken into account
\footnote{As conspicuous below, the calculation amounts to inserting
twice $|\ K_L>_{in}\;_{out}\!<K_L\ | + |\ K_S>_{in}\;_{out}\!<K_S\ | $;
this is justified in the approximation $\epsilon_L = \epsilon_S$
that we use, in which this expression becomes equal to one and where the crossed
scalar products ${_{out}\!<K_L\ |\ K_S>_{in}}$ and ${_{out}\!<K_S\ |\ K_L>_{in}}$
vanish.}
.

Let us calculate the amplitude for a produced $\overline{K^0}$ to decay
into $\ell^+ \nu\pi^-$ after time $(t_f - t_i)$:

\vbox{
\begin{eqnarray}
A(\overline{K^0} \to \ell^+ \nu\pi^-)_{t_f - t_i}  =  {_{out}\!<\ell^+ \nu\pi^-\ |\ 
K_L(t_f)>_{in}}\;{_{out}\!<K_L(t_f)\ |\  K_L(t_i)>_{in}}\;{_{out}\! <K_L(t_i)\
|\  \bar
K^0>} +\cr
 +  {_{out}\!<\ell^+ \nu\pi^- \ |\  K_S(t_f)
>_{in}}\;{_{out}\!<K_S(t_f)\ |\ K_S(t_i)>_{in}}\;{_{out}\!<K_S(t_i)\ |\ \bar
K_0 >}.
\label{eq:CPL1}
\end{eqnarray}
}
Calculating (\ref{eq:CPL1}) we will first neglect the tiny difference between
$\epsilon_L$ and $\epsilon_S$, taking thus
$\epsilon_L^{in} = \epsilon_S^{in} = \epsilon^{in}$.

One uses the expressions for $K_L$ and $K_S$ in (\ref{eq:CPeigeps})
together with the orthogonality relations between $K^0$ and $\overline{K^0}$;
according to the $\Delta Q = \Delta S$ rule,
 only the $K^0$ component of $K_L$ and $K_S$ can produce $\ell^+$; this
yields
\footnote{Time evolution:
having introduced decaying particles fixes the direction of evolution of
time. This is most easily seen by considering the time-dependent
 scalar product of ``in'' and ``out'' mass eigenstates, for example\break
${_{out}<K_L(t_f)\ |\ K_L(t_i)>_{in}}$; if one adopts, like in
\cite{AlvarezGaumeKounnasLolaPavlopoulos}, the ``usual'' time evolution
$|\ K_L(t_i)>_{in} = e^{-iM_L t_i} |\ K_L(0)>_{in}$,
${_{out}<K_L(t_f)\ |} = e^{iM_L t_f} {_{out}<K_L(0)\ |}$,
where $M_L$ and $M_S$ have been defined in (\ref{eq:massdef}), one gets
${_{out}<K_L(t_f)\ |\ K_L(t_i)>_{in}} = e^{-iM_L(t_i-t_f)}
= e^{-im_L(t_i-t_f) -\frac{\Gamma_L}{2}(t_i-t_f)}$, which leads to an
exponential growth since $t_i < t_f$.

The time evolution, which is arbitrary in ordinary QM with a hermitian
Lagrangian (see for example \cite{Landau} paragraphs 6 and 8), becomes here
relevant: the Schr\oe dinger equation has to be chosen here as \hbox{$H\psi =
-i\hbar \frac{\partial\psi}{\partial t}$}, and the time evolution of ``in''
and ``out'' states
\begin{equation}
|\ K_L(t_i)>_{in} = e^{+iM_L t_i} |\ K_L(0)>_{in},\quad
{_{out}<K_L(t_f)\ |} = e^{-iM_L t_f} {_{out}<K_L(0)\ |},
\label{eq:evol1}
\end{equation}
and similar equations for $K_S$.}

\vbox{
\begin{eqnarray}
A(\overline{K^0} \to \ell^+ \nu\pi^-)_{t_f-t_i} & = &
\frac{e^{-i\alpha}}{2D^2}
\left[-\frac{1}{\xi_L^2} e^{-im_L (t_f-t_i) -\frac{\Gamma_L}{2}(t_f-t_i)} +
\frac{1}{\xi_S^2}e^{-i  m_S (t_f-t_i) -\frac{\Gamma_S}{2}(t_f-t_i)} \right] \cr
& \times &  {\cal A}(K^0 \to \ell^+ \nu\pi^-) \;.\cr
&&
\label{eq:CPL2}
\end{eqnarray}
}
Analogously, the amplitude for a produced $K^0$ to decay into $\ell^-
\nu\pi^+$ after time $(t_f-t_i)$ is given by:

\vbox{
\begin{eqnarray}
A(K^0 \to \ell^- \bar\nu\pi^+)_{t_f-t_i} & = &
\frac{e^{i\alpha}}{2D^2}
\left[-\xi_L^2\,e^{-im_L (t_f-t_i) -\frac{\Gamma_L}{2}(t_f-t_i)}
+ \xi_S^2\,e^{-i m_S (t_f-t_i) -\frac{\Gamma_S}{2}(t_f-t_i)} \right]\cr
& \times &  {\cal A}(\overline{K^0} \to \ell^- \nu\pi^+) \;,\cr
&&
\label{eq:CPL3}
\end{eqnarray}
}
and ${\cal A}(\overline{K^0} \to \ell^- \bar\nu\pi^+) =
{\cal A}(K^0 \to \ell^+ \nu \pi^-)$.

In the approximation $\epsilon_L^{in} = \epsilon_S^{in}=\epsilon^{in}$
at which we are working (\ref{eq:nota5}) become
\begin{equation}
\xi_S = \xi_L = \sqrt{\frac{1-\epsilon^{in}}{1+\epsilon^{in}}},
\label{eq:xirhoapp}
\end{equation}
the modulus of which is invariant by the rephasing (\ref{eq:rephase2}).

From (\ref{eq:CPL2}), (\ref{eq:CPL3}) and (\ref{eq:xirhoapp}),
the time dependence and the dependence on
the arbitrary phase $\alpha$ for the $T$-odd asymmetry defined in
(\ref{eq:asyms1}) cancel and we obtain:
\begin{equation}
A_T = \frac{\left|\frac{1+\epsilon^{in}}{1-\epsilon^{in}}\right|^2 -
\left|\frac{1-\epsilon^{in}}{1+\epsilon^{in}}\right|^2}
{\left|\frac{1+\epsilon^{in}}{1-\epsilon^{in}}\right|^2 +
\left|\frac{1-\epsilon^{in}}{1+\epsilon^{in}}\right|^2}
= 4\;\frac{\Re(\epsilon^{in})(1+|\epsilon^{in}|^2)}
{(1+|\epsilon^{in}|^2)^2 + 4 (\Re(\epsilon^{in}))^2} \; .
\label{eq:CPL4}
\end{equation}
This result is independent of the rephasing (\ref{eq:rephase2}), unlike its
approximation by $4\, \Re(\epsilon)$ (see subsection
\ref{subsub:CPparam}) that one finds for
example in \cite{AlvarezGaumeKounnasLolaPavlopoulos}; accordingly, it can
now be evaluated for any value of $\omega$ (or $\Omega_1 (\ref{eq:nota6})$).
However, when $|\epsilon^{in}| \ll 1$, it is well approximated by $4\,
\Re(\epsilon^{in})$.

The corrections that would eventually arise from $\epsilon_L \not=
\epsilon_S$ where checked to vanish by calculating Feynman diagrams as we
do later for $A_{TCP}$ in subsection \ref{subsub:ATCP}. The calculation
makes use of
the explicit form of the non-diagonal $K_L-K_S$ vertex $V$ (\ref{eq:KLKS}),
and of the close approximation to the diagonal term $a(q^2)$
in the inverse Lagrangian (\ref{eq:invpro}) $a(q^2) \approx \frac{1}{2}
\left(q^2-\frac{M_{K_L}^2 + M_{K_S}^2}{2}\right)$.

Another check that we did using the same technique is that each
individual transition amplitude $K^0(t_i) \to \overline{K^0}(t_f)$ and
$\overline{K^0}(t_i) \to K^0(t_f)$ vanishes when $t_f = t_i$ even when
$\epsilon_L \not = \epsilon_S$.

\subsubsection{The semi-leptonic charge asymmetries for $\boldsymbol{K_L}$ and
$\boldsymbol{K_S}$}
\label{subsub:charas}

Supposing that the semi-leptonic decay rates (\ref{eq:dSdQ}) allowed
by the $\Delta S = \Delta Q$ rule  are identical and using
(\ref{eq:CPeigeps}), one gets
\footnote{For example, the transition $K_L \rightarrow \pi^- \ell^+ \nu$
can only occur by a first transition from $K_L$ to $K^0$ and one writes
\hbox{$_{out}\!<\pi^- \ell^+ {\nu}\ |\
K_{L}>_{in} = _{out}\!<\pi^- \ell^+ \overline{\nu}\ |\ K^0><K^0\ |\ K_L>_{in}$}
etc}

\vbox{
\begin{eqnarray}
\delta_L&=&
\frac{\displaystyle\frac{1}{|\xi_L|^2} - |\xi_L|^2}{\displaystyle\frac{1}{|\xi_L|^2} + |\xi_L|^2}
= \frac{|b(z_1)|-|c(z_1)|}{|b(z_1)|+|c(z_1)|}
= \frac{2\Re(\epsilon_L^{in})} {1+|\epsilon_L^{in}|^2},\cr
\delta_S&=&
\frac{\displaystyle\frac{1}{|\xi_S|^2} - |\xi_S|^2}{\displaystyle\frac{1}{|\xi_S|^2} + |\xi_S|^2}
= \frac{|b(z_2)|-|c(z_2)|}{|b(z_2)|+|c(z_2)|}
=\frac{2\Re(\epsilon_S^{in})} {1+|\epsilon_S^{in}|^2}
\label{eq:deltaLS}
\end{eqnarray}
}
and
\begin{equation}
\delta_S - \delta_L=
2\left(\frac{\Re(\epsilon_S^{in})}{1+|\epsilon_{S}^{in}|^2}
 - \frac{\Re(\epsilon_L^{in})} {1+|\epsilon_{L}^{in}|^2}\right).
\label{eq:deltaS-L}
\end{equation}
When $\epsilon_L^{in}=\epsilon_S^{in} = \epsilon^{in}$,
(\ref{eq:deltaS-L}) becomes zero;
$\delta_S - \delta_L$ accordingly measures
the difference between the indirect $CP$ violating parameters
of the $K_L$ and $K_S$ mesons.

This result is to be compared with the one in \cite{Buchanan} and the one
in \cite{RPP}. In \cite{Buchanan}, $TCP$ violation is simply parametrized
by different diagonal elements in the mass matrix; since their $TCP$
violating parameter $\delta$ is precisely defined as the ratio of the
difference of the diagonal elements of the mass matrix and of the difference
of the physical masses, the formula in \cite{RPP}, which comes from
\cite{Buchanan}, is only consistent. In QFT, since $\epsilon_L \not =
\epsilon_S$, this is not a good way to parametrize $TCP$ violation.

$\delta_S$ and $\delta_L$ are both invariant by the rephasing
(\ref{eq:rephase2}); forgetting about their denominators  or approximating
them by $1$ is illegitimate, since then real parts of the $\epsilon$'s is
not invariant by this rephasing (see subsection \ref{subsub:CPparam}).
However, for  $|\epsilon_L^{in}|\ll 1, |\epsilon_S^{in}|\ll 1$, one can
neglect the denominators in (\ref{eq:deltaLS})(\ref{eq:deltaS-L}) and
obtain $\delta_S -\delta_L \approx 2\,\Re(\epsilon_S - \epsilon_L)$.

While $\delta_L$ has already been measured with an average of
$\delta_L = (3.27 \pm 0.12)\ 10^{-3}$ \cite{RPP},
there are only preliminary results for $\delta_S$ coming
from the KLOE detector \cite{KLOE}:
 $\delta_S = (-2 \pm 9_{stat} \pm 6_{syst})\ 10^{-3}$. One is obviously
very far from the precision of order $10^{-17}$ (see subsection
\ref{subsub:deltaeps} below) requested to test the expected difference
between $\delta_L$ and $\delta_S$.

\subsubsection{An estimate of $\boldsymbol{(\epsilon_S^{in} - \epsilon_L^{in})}$}
\label{subsub:deltaeps}

Using the notations of subsection \ref{subsection:QM} and according to the
first line of (\ref{eq:epsilons}) we have:
\begin{equation}
\epsilon_{L,S}^{in} = \frac{1-e^{-i\alpha}\sqrt{\frac{c(z_1,z_2)}{b(z_1, z_2)}}}
{1+e^{-i\alpha}\sqrt{\frac{c(z_1, z_2)}{b(z_1,z_2)}}}
\label{eq:eps3}
\end{equation}
(\ref{eq:eps3}) is expressed only in terms of the non-diagonal elements
$b(z)$ and $c(z)$ of the inverse propagator (\ref{eq:invpro}).
The non-diagonal elements of the QM mass matrix (\ref{eq:MMQ}) being
expected to be close to the ones of the inverse propagator, and since we
only need an order of magnitude estimate of the
difference $\epsilon_S - \epsilon_L$, we shall approximate (\ref{eq:eps3})
by
\begin{equation}
\epsilon_{L,S}^{in}
\approx \frac{1-e^{-i\alpha}\sqrt{\frac{\overline{m_{12}(z_1,z_2)}
-\frac{i}{2}\overline{\gamma_{12}}(z_1,z_2)}{m_{12}(z_1,z_2)
-\frac{i}{2}\gamma_{12}(z_1,z_2)}}}
{1+e^{-i\alpha}\sqrt{\frac{\overline{m_{12}(z_1,z_2)}
-\frac{i}{2}\overline{\gamma_{12}}(z_1,z_2)}{m_{12}(z_1,z_2)
-\frac{i}{2}\gamma_{12}(z_1,z_2)}}},
\label{eq:eps3bis}
\end{equation}

where $m_{12}$ and $\gamma_{12}$ should be taken at $q_1^2 = z_1 =
m_L^2$ and $q_2^2 = z_2 = m_S^2$ for $\epsilon_L$ and
$\epsilon_S$, correspondingly
\footnote{The mass matrix (\ref{eq:MMQ}) of QM has dimension
$[mass]$ while the inverse propagator (\ref{eq:invpro}) of QFT
has dimension $[mass]^2$. Nevertheless, in (\ref{eq:eps3bis}), the ratio
of the non-diagonal
elements $b$ and $c$ of the inverse propagator has been identified with the
ratio of the non-diagonal elements of the quantum mechanical mass matrix.
This is  a good enough approximation for the order of
magnitude estimate that we want to get as soon as the
mass matrix in QFT is recognized to be the square of the mass matrix in QM
and its non-diagonal elements are much smaller than its diagonal elements.}
.
The standard choice $\alpha=0$ for the arbitrary
phase $\alpha$ corresponds to the condition $\gamma_{12} =
\overline{\gamma_{12}}$ (in the quark model this choice is equivalent to that
of real $V_{us}$ and $V_{ud}$ CKM matrix elements):
\begin{equation}
\epsilon_{L,S}^{in} = \frac{\sqrt{m_{12}-\frac{i}{2}\gamma_{12}} -
{\sqrt{\overline{m_{12}}-\frac{i}{2}\gamma_{12}}}}{\sqrt{m_{12}-\frac{i}{2}\gamma_{12}}
+ {\sqrt{\overline{m_{12}}-\frac{i}{2}\gamma_{12}}}}.
\label{eq:eps4}
\end{equation}
To proceed with our estimate, we shall hereafter rely on the quark picture
of neutral mesons, and the so-called ``box diagrams'' which generate
$K^0-\overline{K^0}$ transitions
\cite{Vysotsky}\cite{BrancoLavouraSilva}\cite{Belusevic}\cite{BigiSanda}.

$m_{12}$ is almost real; a nonzero phase is generated by the 
quark box diagram with $t\bar t$ exchange which is highly suppressed
by the smallness of the CKM matrix elements $V_{ts}$ and $V_{td}$
\cite{RPP}. One has \cite{Vysotsky} $m_{12} \approx \Re(m_{12})
 \approx \gamma_{12} \gg \Im(m_{12})$, and, expanding (\ref{eq:eps4}),
one gets
\begin{equation}
\epsilon_{L,S}^{in} \approx \frac{i\, {\Im}(m_{12})}{2(m_{12}
-\frac{i}{2}\gamma_{12})}
\approx \frac{i\, {\Im}(m_{12})}{m_S - m_L
-\frac{i}{2}(\Gamma_S - \Gamma_L)} = \frac{-i\,
{\Im}(m_{12})}{\Delta m_{LS} + \frac{i}{2} \Gamma_S},
\label{eq:est3}
\end{equation}
where we have defined
\begin{equation}
M_L = m_L -i\frac{\Gamma_L}{2}, M_S = m_S -i\frac{\Gamma_S}{2},
\label{eq:massdef}
\end{equation}
neglected $\Gamma_L$ in comparison with $\Gamma_S$ and
substituted $\Delta m_{LS} \equiv m_L - m_S$
\footnote{The relation between the first and the second denominators of
(\ref{eq:est3}) is obtained from the expression of the eigenvalues of
(\ref{eq:MMQ}) by using $\gamma_{11}\ll
m_{11}$, $\gamma_{22}\ll m_{22}$, $m_{12}$ quasi real, and choosing the
phase convention such that $\gamma_{12} = \overline{\gamma_{12}}$.}
.

There exists a  dependence of $m_{12}$  and $\gamma_{12}$ on the momentum
$q^2$, which leads to a tiny difference between $\epsilon_L$ and
$\epsilon_S$.

The dominant contribution to ${\Im}(m_{12})$ is produced by the box diagram
with two intermediate $t$-quarks ($G_F$ is the Fermi constant):
\begin{equation}
[{\Im}(m_{12})]_{tt} \propto \lambda^{10} G_F^2(m_t^2 + q^2)\; ,
\label{eq:est4}
\end{equation}
where $\lambda$ is the Cabibbo angle. However, the sub-dominant diagram
with two intermediate $c$-quarks generates a larger $q^2$ dependence
(it contributes to ${\Im}(m_{12})$ since ${\Im}(V_{cd}) \sim
\lambda^5$):
\begin{equation}
[{\Im}(m_{12})]_{cc} \propto \lambda^6 G_F^2 (m_c^2 + q^2) \; .
\label{eq:est5}
\end{equation}
As a result, since $\lambda^4 m_t^2 \gg m_c^2$, one gets:
\begin{equation}
{\Im}(m_{12})(q^2) \approx {\Im}(m_{12})(0)\left(1+\frac{q^2}{\lambda^4
m_t^2}\right)\; .
\label{eq:est6}
\end{equation}

Since $\Delta m_{LS}$ is numerically close to $\Gamma_S/2$, the
contributions of $c$ and $u$ quarks to $m_{12}$ should be comparable. In
this way, we get
\begin{equation}
m_{12}(q^2) \sim \lambda^2(m_c^2 +q^2) \approx m_{12}(0)
\left(1+ {\cal O}(\frac{q^2}{m_c^2})\right) \;,
\label{eq:est7}
\end{equation}
where the contribution of the box diagrams with intermediate
$c$ and $u$ quarks is taken into account.

Finally, the $q^2$ dependence of $\epsilon_{L,S}$ is determined by that of
 $\gamma_{12}$; the dependence of $\gamma_{12}$ originates
 both from $K \to \pi\pi$ matrix elements and two-pions phase space
\begin{equation}
\gamma_{12}(q^2) \approx \gamma_{12}(0)
\left(1+ {\cal O}(\frac{q^2}{m_K^2})\right).
\label{eq:est8}
\end{equation}
Taking into account the dominant contributions (\ref{eq:est7}) (\ref{eq:est8})
we obtain
\begin{equation}
\epsilon_L^{in} - \epsilon_S^{in} \sim
\epsilon \frac{\Delta m_{LS}}{m_K} \sim 10^{-17}.
\label{eq:estimeps}
\end{equation}
In this way we obtain that the $q^2$ dependence of the kaon self-energy
leads to different values of $\epsilon_L$ and $\epsilon_S$: 
the statement that this difference signals $TCP$ violation is seen to be
wrong.

\subsubsection{The asymmetry $\boldsymbol{A_{TCP}}$}
\label{subsub:ATCP}

We have seen that finding a non vanishing difference of charge
asymmetries $\delta_S - \delta_L$ cannot be {\em a priori}
interpreted as a signal of $TCP$ non-invariance, because this
difference is allowed to keep non-vanishing even when $TCP$ is
achieved. If it exceeds an expected $10^{-17}$ (see (\ref{eq:estimeps})),
one clearly has a problem.

The question arises whether some observable should identically
vanish when $TCP$ symmetry holds. We show below that it is the case of
$A_{TCP}$ asymmetry given by the third equation of (\ref{eq:asyms1}).
$A_{TCP}$ is analogous to $A_T$, but,
this time, the asymmetry for ``allowed'' semi-leptonic decays is studied.

We explicitly calculate $K^0 \rightarrow K^0$ and $\overline{K^0} \rightarrow
\overline{K^0}$ transitions in QFT by calculating the corresponding Feynman
diagrams, allowing $\epsilon_L$ to be different from $\epsilon_S$ and
$K_L \leftrightarrow K_S$ transitions.

The diagrams that we evaluate are, for $K^0 \rightarrow K^0$ transitions,
drawn in Fig.~1. The same type of diagrams occur, of course, for
$\overline{K^0} \rightarrow \overline{K^0}$ transitions. We have drawn diagrams only up
to first order in the $K_L-K_S$ coupling $V$,
but results have been checked to stay unchanged at second order in $V$.
\figskip

\vbox{
\begin{center}
\includegraphics[height=3truecm,width=14truecm]{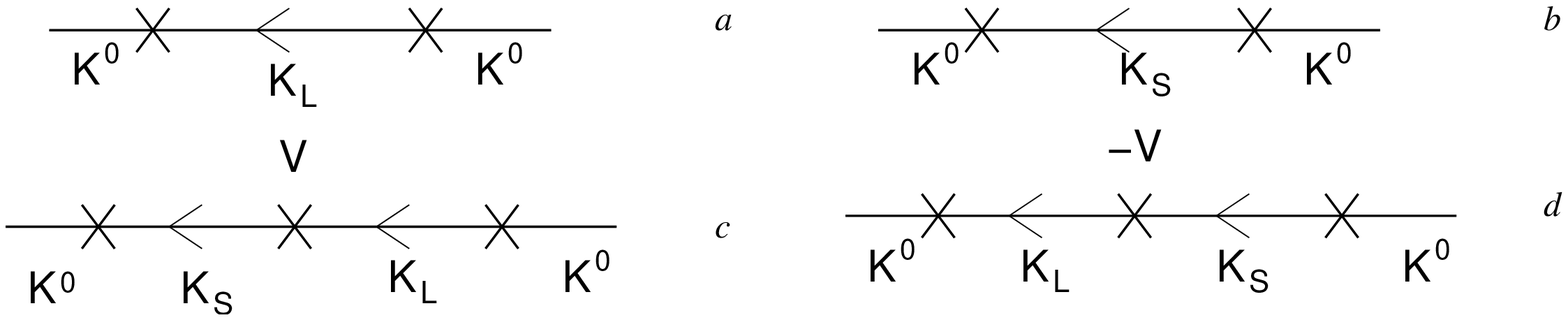}
\end{center}
\figskip
\centerline{\em Fig.~1: Feynman diagrams for $K^0$ to $K^0$ transition up
to first order in $V$.}
\figskip
}

$\bullet$ {\bf The $\boldsymbol{K_L-K_S}$ couplings.}

While transitions between $K_L$ and $\tilde{K_S}$, or $K_S$ and
$\tilde{K_L}$ are forbidden, the ones between $K_L$ and $K_S$ are
authorized, as shown by a very simple calculation.

(\ref{eq:invpro}) is equivalent to writing the $TCP$ invariant Lagrangian
${\cal L}_{TCP}$ in the $(K^0, \overline{K^0})$ basis:
\begin{equation}
{\cal L}_{TCP}(z) = a(z)\left( |\ K^0><K^0\ | + |\
\overline{K^0)}><\overline{K^0}\
|\right) -b(z) |\ K^0><\overline{K^0}\ | -c(z) |\ \overline{K^0}><K^0\ |.
\label{eq:Ltcp}
\end{equation}
Using then (\ref{eq:flaveigeps}) to express $K^0$ and $\overline{K^0}$ in terms
of $K_L$ and $K_S$, one obtains the $K_L-K_S$ couplings
\begin{equation}
{\cal L}_{TCP}(z) \ni \frac{1}{2D^2} \left(
a(z)\left(\frac{\xi_L}{\xi_S} - \frac{\xi_S}{\xi_L}\right)
+ b(z) e^{i\alpha}\xi_L\xi_S -c(z)e^{-i\alpha}\frac{1}{\xi_L\xi_S}\right)
\bigg[ {|\ K_S>_{in}}\,{_{out}\!<K_L\ |} - {|\ K_L>_{in}}\,{_{out}\!<K_S\
|}\bigg].
\label{eq:KLKS}
\end{equation}
The important point for our concern is that, at any $z=q^2$, the 
$K_L$ to $K_S$ coupling $V(z)$ is the opposite of the $K_S$ to $K_L$
coupling.

$\bullet$ {\bf Explicit calculation}

Let us calculate explicitly the diagram c of Fig.~1, evaluated from
right to left.

- The first vertex is the projection of $K^0$ on $K_L$, {\em i.e.} the scalar
product ${_{out}\!<K_L\ |K^0>}$; it is, according to  (\ref{eq:CPeigeps}),
$\xi_L/\sqrt{2}$;

- the $K_L$ propagator $1/(q^2 - M_L^2)$ follows;

- the second vertex is the $K_L$ to $K_S$ vertex $V(q^2)$;

- the $K_S$ propagator $1/(q^2 - M_S^2)$ follows;

- the last vertex is the projection of $K_S$ on $K^0$, that is the scalar
  product $<K^0\ |\ K_S>_{in}$; it is given by (\ref{eq:CPeigeps}) and is
equal to $1/(\sqrt{2}\xi_S)$.

Finally, the diagram c of Fig.~1 is given by
$\frac{V(q^2)}{2}\frac{\xi_L}{\xi_S}\frac{1}{q^2-M_L^2}\frac{1}{q^2-M_S^2}$.

So doing for all diagrams, one gets for the amplitudes ${\cal A}_{K^0
\rightarrow K^0}$ and ${\cal A}_{\overline{K^0}\rightarrow \overline{K^0}}$
\begin{equation}
{\cal A}(q^2)_{K^0\rightarrow K^0}=
\frac{1}{2}\left(
\frac{1}{q^2-M_L^2} + \frac{1}{q^2-M_S^2}
+ \frac{V(q^2)}{2}\left(\frac{\xi_L}{\xi_S} - \frac{\xi_S}{\xi_L}\right)
\frac{1}{q^2-M_L^2}\frac{1}{q^2-M_S^2}\right)
={\cal A}(q^2)_{\overline{K^0}\rightarrow \overline{K^0}}.
\label{eq:ampli}
\end{equation}
The Fourier transforms of these two amplitudes are identical,
too. Their $t$ dependence is just the dependence of the corresponding
$<K^0(t_f)\ |\ K^0(t_i)>$ and
 $<\overline{K^0}(t_f)\ |\ \overline{K^0}(t_i)>$ on $t_f - t_i$.

Then, according to the third equation in (\ref{eq:asyms1}), 
\begin{equation}
A_{TCP} = 0.
\label{eq:Atcp}
\end{equation}
This result is to be compared with \cite{RPP}, where $A_{TCP}$ is mentioned
 to be equal to $\delta_S - \delta_L$. We have shown that this is not the
case: despite $\epsilon_L \not = \epsilon_S$, $A_{TCP}$ vanishes when $TCP$
is realized. It is easy to trace the root of this mechanism in
(\ref{eq:KLKS}) valid for a $TCP$ invariant Lagrangian.

\subsection{Testing $\boldsymbol{TCP}$?}
\label{subsection:test?}

Testing an eventual violation of $TCP$ in binary systems of neutral mesons
becomes a more and more important concern for both theorists
\cite{Okun1}\cite{Okun3} \cite{Pavlopoulos}
and experimentalists
\cite{BriereOrr} \cite{Schwingenheuer} \cite{Alavi-Harati} \cite{CPLEAR} \cite{BaBar} \cite{Buchanan}.

The measurement of $\delta_S - \delta_L$  was proved in
this work not to be a test of $TCP$. If it is detected to exceed the
estimated value (\ref{eq:estimeps})
\footnote{Remember that this estimate was done with a precise phase
convention.}
,  questions would arise, but the last
should most probably be whether $TCP$ symmetry is broken. It is useful to
recall that our calculations have been performed  supposing the $\Delta S
= \Delta Q$ rule exactly satisfied (see footnote \ref{footnote:dSdQ}),
assumption which could of course need a revision.

The non-vanishing of $A_{TCP}$ stays, at the opposite, a clean test of a
violation of $TCP$.

Other experimental signals like the detection of a slight amount of
 longitudinal polarization
for the emitted photons  in $\pi^0 \rightarrow \gamma \gamma$ decays
 \cite{Okun3} could provide a test of $TCP$, according to their
description in the usual framework of a local field theory.

The subject of investigating all possible tests of $TCP$ violation goes
anyhow beyond the scope of this work \cite{Okun3}
 and we shall not comment more on this subject here.

One should also always keep in mind that the logic of introducing explicit $TCP$
violating parameters in a local field theory can appear questionable since
the latter presupposes the former.

\section{CONCLUSION}
\label{section:conclusion}

This work is a succession of elementary deductions from basic
properties of propagators in QFT. The results are simple
and unambiguous.

The main results of the paper can be summarized as follows.
Taking the example of neutral kaons, we exhibited substantial differences
between the treatments of binary
systems of neutral mesons in QM and in QFT.
The role of the $TCP$ symmetry has been clarified, and QM has been shown to
yield an improper characterization of this symmetry. An essential role is
played by the definition of the physical masses as the poles of the
full propagator; this smooths out conceptual problems linked with the
existence of ``in'' and ``out'' eigenstates, and 
predicts a difference between the $CP$ violating
parameters of $K_L$ and $K_S$, which originates from their mass splitting
 and  is not a characteristic signal of $TCP$ violation.
While the asymmetry $A_{TCP}$ stays nevertheless a good test of the $TCP$
symmetry, $\delta_S - \delta_L$  has been  shown
 to test the difference $\epsilon_S - \epsilon_L$; the latter can be
different from zero even when $TCP$ is realized..

We have shown that QM formul\ae\, often quoted in the literature depend on
the arbitrary rephasing of $K^0$ and $\overline{K^0}$, and are, hence, not
physically relevant; we have given the correct, phase independent,
formul\ae\, obtained in QFT.

 The introduction of a (constant) unique mass matrix to
describe these binary systems is inappropriate.
The correct way to diagonalize a general complex propagator is by
using a bi-orthogonal basis and not  a bi-unitary transformation.
Finally, local QFT, which presupposes $TCP$ symmetry, is not
an appropriate framework to parameterize $TCP$ violation.

A similar study will be devoted to fermions
\cite{MachetNovikovVysotsky} with a special emphasis on mixing matrices
and unitarity.

\vskip .75cm

\underline{{\em Acknowledgments}}

\medskip
{\em A very special thank is due to R. Stora for his invaluable
help. Sections \ref{subsection:anacons} and \ref{subsection:discons} are
strongly inspired from his letters.
The authors are very indebted to D.V.~Ahluwalia, L.~Alvarez-Gaum\'e,
O.~Babelon, M.~Bellon, L.B.~Okun, and J.B.~Zuber
for many discussions,  exchanges and suggestions; this work owes a lot to
their cleverness and their patience. They also thank Ya.I.~Azimov
for pointing their attention to the very relevant work \cite{Azimov} and
useful email exchanges. \newline\null
B.M. wants to thank the lab.\,180 of ITEP for the very warm and friendly
hospitality extended to him on several occasions. A large part of this work
was conceived there.\newline\null
M.V. wants to thank LPTHE and University of Paris 7 for the possibility
offered to him to spend April 2004 and to work there.\newline\null
V.A.\,N. and M.I.\,V. are supported by RFBR grant $\sharp$ 00-15-96562 and by NTP FYaF
40.052.1.1.1112.
}

\newpage
%
\appendix
%

\section{CONSTRAINTS SET BY DISCRETE SYMMETRIES AND LORENTZ INVARIANCE ON
THE PROPAGATOR OF NEUTRAL KAONS}
\label{section:disconst}

We work in the $(K^0,\overline{K^0})$ basis.

Let $\varphi_{K^0}(x)$ be the Heisenberg operator for $K^0$ at space time point
$x = (\vec x,t)$ and $\varphi_{K^0}(\vec x)$ the corresponding
Schr\oe dinger operator (see also subsection \ref{subsub:CPanti}).
Since other fields will be related to it, we shall often omit the
corresponding subscript, writing instead $\varphi$ when it is the only one
appearing in a formula.

\subsection{DEFINITION OF $\boldsymbol{C}$, $\boldsymbol{P}$ and
$\boldsymbol{CP}$ TRANSFORMATIONS
ON FIELD OPERATORS}
\label{subsection:action}

The action of these symmetry transformations is defined for the
Schr\oe dinger fields.

\subsubsection{Charge conjugation $\boldsymbol C$}
\label{subsub:chacon}

The charge conjugation operation $C$ transforms a multiplet of an internal
symmetry into a multiplet of the complex conjugate representation; we thus
define, for field operators
\begin{equation}
C \varphi_{K^0}(\vec x) C^{-1} =
                   e^{-i\alpha} \varphi_{\overline{K^0}}(\vec x),
\label{eq:Cop}
\end{equation}
where we have introduced an arbitrary phase $\alpha$; then,
$\varphi_{\overline{K^0}}$ and $\varphi^\dagger_{K^0}$ are again connected by
an arbitrary phase $\delta$
\begin{equation}
\varphi_{\overline{K^0}}(\vec x)= e^{-i\delta}\varphi^\dagger_{K^0}(\vec x).
\label{eq:bardag}
\end{equation}
such that
\begin{equation}
C \varphi_{K^0}(\vec x) C^{-1} =
                   e^{-i(\alpha+\delta)} \varphi^\dagger_{K^0}(\vec x).
\label{eq:Cop2}
\end{equation}
$C$ is a unitary operator $CC^\dagger = 1 = C^\dagger C$, which entails
that $C^\dagger = C^{-1}$.

We find accordingly
$$C\varphi^\dagger_{K^0}(\vec x)C^{-1}
= C\varphi^\dagger_{K^0}(\vec x)C^\dagger
= (C \varphi_{K^0}(\vec x)C^\dagger)^\dagger = e^{i(\alpha + \delta)}
\varphi_{K^0}(\vec x),$$
\footnote{In the second equality we have only replaced $C^{-1}$ by
$C^\dagger$.\label{footnote:C}}
and
$$C\varphi_{\overline{K^0}}(\vec x)C^{-1} = C e^{-i\delta}
\varphi^\dagger_{K^0}(\vec x)C^{-1} = e^{-i\delta}
C \varphi^\dagger_{K^0}(\vec x)C^{-1} =
e^{-i\delta}e^{i(\alpha + \delta)}
\varphi_{K^0}(\vec x) = e^{i\alpha}\varphi_{K_0}(\vec x).$$

It can then immediately be checked that
\begin{equation}
C^2 = 1.
\label{eq:Ccarre}
\end{equation}

\subsubsection{Parity $\boldsymbol P$}
\label{subsub:pari}

The parity operator $P$ is also a unitary operator $PP^\dagger=1=P^\dagger
P$; allowing again for an arbitrary phase $\beta$, it acts on field
operators according to
\begin{equation}
P \varphi_{K^0}(\vec x) P^{-1} = e^{i\beta} \varphi_{K^0}(-\vec x).
\label{eq:Pop}
\end{equation}
One has then
\begin{equation}
P \varphi_{K^0}^\dagger(\vec x) P^{-1}
= e^{-i\beta} \varphi_{K^0}^\dagger(-\vec x);
\label{eq:Popstar}
\end{equation}
indeed: $P \varphi_{K^0}^\dagger(\vec x) P^{-1}
= P \varphi_{K^0}^\dagger(\vec x) P^\dagger
= \left(P \varphi_{K^0}(\vec x) P^\dagger\right)^\dagger$
\footnote{See footnote \ref{footnote:C}.}
.

Replacing in (\ref{eq:Popstar}) $\varphi_{K^0}^\dagger(\vec x)$ by
$e^{i\delta}\varphi_{\overline{K^0}}(\vec x)$ according to (\ref{eq:Cop})
yields
\begin{equation}
P\varphi_{\overline{K^0}}(\vec x)P^{-1} =
e^{-i\beta}\varphi_{\overline{K^0}}(-\vec x).
\label{eq:Pop2}
\end{equation}
Notice that the product of the phases occurring in (\ref{eq:Pop}) and
(\ref{eq:Pop2}) is $+1$: the relative intrinsic parity of
$\varphi_{K^0}$ and $\varphi_{\overline{K^0}}$ (or $\varphi^\dagger_{K^0}$) is
$+1$.

The operation of parity transformation is a geometrical operation which,
repeated twice, should give the identity; one can accordingly impose
\begin{equation}
P^2=1.
\label{eq:Pcarre}
\end{equation}
This entails that $e^{2i\beta} = 1$; since kaons are pseudoscalar, we
will choose hereafter
\begin{equation}
e^{i\beta}=-1.
\label{eq:pseudo}
\end{equation}
\subsubsection{$\boldsymbol{CP}$ transformation}
\label{subsub:chapar}

We then find the laws of transformation by the combined symmetry $CP$;
it is instructive that requesting that the two operators $C$ and $P$
commute (or anticommute), $[C,P]=0$ ($\{C,P\}=0$) yields the same condition
on the phase $\beta$ of the parity transformation as the one given by $P^2 =1$
\footnote{That $C$ and $P$ commute is indeed not the only possible choice.
They can also anticommute, which leads in particular to the so-called
Wigner bosons \cite{Ahluwalia}.}
. We calculate in two
ways $CP \varphi_{K^0}(\vec x) (CP)^{-1}$, using the linearity of both
operators:

$CP \varphi_{K^0}(\vec x) (CP)^{-1} = C(P \varphi_{K^0}(\vec x)
P^{-1})C^{-1}
= C(e^{i\beta} \varphi_{K^0}(-\vec x))C^{-1}
= e^{i\beta} C \varphi_{K^0}(-\vec x)C^{-1}\newline\null
= e^{i(\beta-\alpha)} \varphi_{\overline{K^0}}(-\vec x)
= e^{i(\beta-\alpha-\delta)} \varphi^\dagger_{K^0}(-\vec x)$;

 and

$CP \varphi_{K^0}(\vec x) (CP)^{-1} = P(C \varphi_{K^0}(\vec x)
C^{-1})P^{-1}
= P(e^{-i\alpha} \varphi_{\overline{K^0}}(\vec x))P^{-1}
= e^{-i\alpha} P \varphi_{\overline{K^0}}(\vec x)P^{-1}\newline\null
= e^{-i(\alpha+\beta)} \varphi_{\overline{K^0}}(-\vec x)
= e^{-i(\alpha+\beta+\delta)} \varphi^\dagger_{K^0}(-\vec x)$.

For these two expressions to be identical, one needs
$e^{-i\beta} = e^{i\beta}$, which is the condition obtained in the previous
subsection. With our choice (\ref{eq:pseudo}), we get
\begin{equation}
CP \varphi_{K^0}(\vec x) (CP)^{-1} =
-e^{-i\alpha}\varphi_{\overline{K^0}}(-\vec x) = - e^{-i(\alpha+\delta)}
\varphi^\dagger_{K^0}(-\vec x).
\label{eq:CPop}
\end{equation}
One then gets
$$CP \varphi^\dagger_{K^0}(\vec x) (CP)^{-1}
= CP \varphi^\dagger_{K^0}(\vec x) (CP)^\dagger
= (CP \varphi_{K^0}(\vec x) (CP)^\dagger)^\dagger
= -e^{i(\alpha + \delta)} \varphi_{K^0}(-\vec x),$$
and

\begin{equation}
CP \varphi_{\overline{K^0}}(\vec x) (CP)^{-1} =
-e^{i\alpha}\varphi_{K^0}(-\vec x).
\label{eq:CPop1}
\end{equation}
One has $(CP)^2 =1$.

\subsection{THE NEUTRAL KAON PROPAGATOR}
\label{subsection:propagator}

The notations and definition of the neutral kaon propagator have been given
in subsection \ref{subsection:definition}. We do not repeat them here.

\subsection{CONSTRAINTS SET BY $\boldsymbol{CP}$ SYMMETRY ON THE PROPAGATOR OF
NEUTRAL KAONS}
\label{subsection:CPcons}

\subsubsection{Constraint set by $\boldsymbol{CP}$ symmetry on the anti-diagonal
elements}
\label{subsub:CPanti}

Using (\ref{eq:CPop}), let us investigate the consequences of $CP$
invariance on the propagator.
Since $CP$ is unitary, the following is an identity

\vbox{
\begin{eqnarray}
-g(\vec x,t)&=&e^{i\delta} <CP\ 0 \ |\ \vartheta(t)
CP\varphi(\frac{\vec x}{2},\frac{t}{2}) (CP)^{-1}
CP\varphi(-\frac{\vec x}{2},-\frac{t}{2}) (CP)^{-1} \cr
&&+ \vartheta(-t) CP\varphi(-\frac{\vec x}{2},-\frac{t}{2})(CP)^{-1}
CP\varphi(\frac{\vec x}{2},\frac{t}{2})CP^{-1}\ |\ CP\ 0>.
\end{eqnarray}
}
The theory is invariant by $CP$ if and only if $[CP,H]=0$,
$H$ being the Hamiltonian.
The Heisenberg field $\varphi(\frac{\vec x}{2},\frac{t}{2})$ can be
expressed in term of the Schr\oe dinger field $\varphi(\frac{\vec x}{2})$ by
\begin{equation}
\varphi(\frac{\vec x}{2},\frac{t}{2}) =
e^{i\frac{t}{2}H} \varphi(\frac{\vec
x}{2}) e^{-i\frac{t}{2}H},
\end{equation}
such that $CP$ invariance yields
\begin{equation}CP\varphi_{K^0}(\frac{\vec x}{2},\frac{t}{2})(CP)^{-1}
= e^{i\frac{t}{2}H} CP \varphi_{K^0}
(\frac{\vec x}{2})(CP)^{-1}e^{-i\frac{t}{2}H},
\end{equation}
and, using (\ref{eq:CPop}),
\begin{equation}
CP\varphi_{K^0}(\frac{\vec x}{2},\frac{t}{2})(CP)^{-1}
 = -e^{-i(\alpha+\delta)}\varphi^\dagger_{K^0}(-\frac{\vec x}{2},\frac{t}{2});
\label{eq:CPop2}
\end{equation}
the Heisenberg field transforms in the same way as the Schr\oe dinger field.

Using the invariance of the vacuum and (\ref{eq:CPop2}),
the starting identity becomes
\begin{eqnarray}
-g(\vec x,t)&\stackrel{CP}{=}&
e^{i\delta} <0\ |\ \vartheta(t) (-)e^{-i(\alpha+\delta)}
\varphi^\dagger(-\frac{\vec x}{2},\frac{t}{2})
(-)e^{-i(\alpha+\delta)}\varphi^\dagger(\frac{\vec x}{2},-\frac{t}{2})\cr
&+& \vartheta(-t)(-)e^{-i(\alpha+\delta)}
\varphi^\dagger(\frac{\vec x}{2},-\frac{t}{2})
(-)e^{-i(\alpha+\delta)}\varphi^\dagger(-\frac{\vec x}{2},\frac{t}{2})\ |\ 0>
= e^{-2i\alpha}(-h(-\vec x,t)).\cr
&&
\end{eqnarray}
Lorentz invariance of the propagator (\ref{eq:Lorentz}),
 in the sense discussed in subsection \ref{subsection:definition},
entails $h(-\vec x,t) = h(\vec x,t)$, and one gets
\begin{equation}
g(\vec x,t) \stackrel{CP}{=} e^{-2i\alpha} h(\vec x,t).
\label{eq:CPcons1}
\end{equation}

\subsubsection{Constraint set by $\boldsymbol{CP}$ symmetry on the diagonal elements}
\label{subsub:CPdia}

Going along similar lines, one gets:
\begin{eqnarray}
CP\ d(\vec x,t)
&=& <0\ |\ \vartheta(t)(-)e^{-i(\alpha+\delta)}
\varphi^\dagger(-\frac{\vec x}{2},\frac{t}{2})
(-)e^{i(\alpha+\delta)}\varphi(\frac{\vec x}{2}, -\frac{t}{2})\cr
&+& \vartheta(-t)
(-)e^{i(\alpha+\delta)}\varphi(\frac{\vec x}{2}, -\frac{t}{2})
(-)e^{-i(\alpha+\delta)}\varphi^\dagger(-\frac{\vec x}{2},\frac{t}{2}) \ |\ 0>
= f(-\vec x,t).
\end{eqnarray}
The phases cancel, and
Lorentz invariance of the propagator (\ref{eq:Lorentz}),
 in the sense discussed in subsection \ref{subsection:definition},
entails $f(-\vec x,t) = f(\vec x,t)$; finally we obtain
\begin{equation}
CP d(\vec x,t) = f(\vec x,t).
\label{eq:CPcons2}
\end{equation}
A $CP$ transformations swaps the diagonal elements; they are accordingly
identical if $CP$ invariance is satisfied.

\subsubsection{Final remarks on $\boldsymbol{CP}$}
\label{subsub:concluCP}

The diagonal elements of a $CP$ invariant kaon propagator are identical,
and its anti-diagonal elements are identical up to a phase $\alpha$.

Accordingly, a $CP$ invariant kaon propagator is always normal;
it is equivalent to saying that a non-normal propagator cannot describe
a $CP$ invariant theory; but this leaves the freedom for a normal mass
matrix to also accommodate for $CP$ violation
\footnote{This because $(A \Rightarrow B)$ entails $(\not\!\! B \Rightarrow
\not\!\! A)$, but does not entail $B \Rightarrow A$, which is a wrong
statement: if $B$ is true, $A$ can be either true or false.}
. This is emphasized in the core of the paper.

Notice that proper Lorentz invariance enabled us to transform the $(-\vec
x,t)$ dependence in the propagator into a $(\vec x,t)$ dependence without
making any hypothesis concerning the link between $\varphi(-\vec x,t)$ and
$\varphi(\vec x,t)$.

\subsection{THE WIGHTMAN CONVENTION FOR $\boldsymbol{TCP}$ TRANSFORMATION
\cite{StreaterWightman}\cite{Lee2}\cite{Haag}\cite{BogolubovLogunovOksakTodorov}
\cite{ItzyksonZuber}\cite{Weinberg}\cite{Greenberg}
\cite{BrancoLavouraSilva}\cite{Sachs}}
\label{subsection:Wightman}

The $TCP$ transformation $\Theta$ exists independently of the three individual
transformations $P$, $C$ and $T$.

One defines it on Heisenberg fields because it also concerns  time
evolution for operators and eigenstates.

While phases can appear in the individual transformations $P$, $C$ (and
$T$), there is no arbitrary phase in  $\Theta$.
\cite{StreaterWightman}.

$\Theta$ satisfies
\begin{equation}
\Theta = \Theta^{-1}.
\label{eq:invtheta}
\end{equation}
It is an antiunitary  operator:
\begin{equation}
<\Theta A\ |\ \Theta B> = <A\ |\ B>^\ast = <B\ |\ A>.
\label{eq:antiunit1}
\end{equation}
One deduces in particular from (\ref{eq:antiunit1}), for any operator
${\cal O}(\vec x,t)$
\begin{equation}
<\Theta A\ |\  \Theta {\cal O}(\vec x,t)\Theta^{-1}\ |\ \Theta B>
= <B\ |\  {\cal O}^\dagger(\vec x,t)\ |\ A>.
\label{eq:antiunit2}
\end{equation}
Indeed:
$<\Theta A\ |\  \Theta {\cal O}(\vec x,t)\Theta^{-1}\ |\ \Theta B>
= <\Theta A\ |\  \Theta {\cal O}(\vec x,t)\ |\ B> = <\Theta A\ |\ \Theta({\cal O}(\vec
x,t)B)> \newline\null
\stackrel{(\ref{eq:antiunit1})}{=}
<{\cal O}(\vec x,t)B\ |\ A> = <B\ |\  {\cal O}^\dagger(\vec x,t)\ |\ A>.$

$\Theta$ is an antilinear operator: it complex conjugates all c-numbers on
its right
\begin{equation}
\Theta (a \ |\ A>) = a^\ast \Theta \ |\ A>.
\label{eq:antilin}
\end{equation}
(\ref{eq:antilin}) can easily be obtained from (\ref{eq:antiunit2}) by
replacing the operator $\cal O$ by the c-number $a$.

By a $TCP$ transformation,
in addition to the 4-inversion $(\vec x,t) \rightarrow
(-\vec x, -t)$, any operator should be changed into its hermitian conjugate.
\begin{equation}
\Theta \varphi(\vec x, t)\Theta^{-1} = \varphi^\dagger(-\vec x,-t).
\label{eq:WitLee}
\end{equation}
In this work, we consider the existence of  an antiunitary operator
$\Theta$ satisfying (\ref{eq:WitLee}) as the criterion for $TCP$ invariance.

\subsubsection{Constraint linked to $\boldsymbol{TCP}$ symmetry on the diagonal
elements}
\label{subsub:TCPdiaWL}

By definition (see (\ref{eq:prop}))
\begin{eqnarray}
d(\vec x,t) &=& <{K^0}\ |\ \Delta(\vec x,t)\ |\ K^0>
=<0\ |\ T\{\varphi(\frac{\vec x}{2},\frac{t}{2})
\varphi^\dagger(-\frac{\vec x}{2}, -\frac{t}{2})\}\ |\ 0>\cr
&=& <0\ |\ \vartheta(t)\varphi(\frac{\vec x}{2},\frac{t}{2})
\varphi^\dagger(-\frac{\vec x}{2}, -\frac{t}{2})
+ \vartheta(-t)\varphi^\dagger(-\frac{\vec x}{2}, -\frac{t}{2})
\varphi(\frac{\vec x}{2},\frac{t}{2}) \ |\ 0>.
\label{eq:d0}
\end{eqnarray}
The vacuum is supposed to be unique and invariant by $TCP$: $\Theta\  |0> =
|\ 0>$; one replaces accordingly $|\ 0>$ by $|\ \Theta\ 0>$; one then uses
 (\ref{eq:WitLee}) to replace $\varphi(\frac{\vec x}{2},\frac{t}{2})$
by $\Theta^{-1} \varphi^\dagger(-\frac{\vec x}{2},-\frac{t}{2}) \Theta$,
which is identical to
$\Theta \varphi^\dagger(-\frac{\vec x}{2},-\frac{t}{2}) \Theta^{-1}$
because of (\ref{eq:invtheta}), and  one gets
\footnote{One does the same for the second operator}
\begin{eqnarray}
d(\vec x,t) &=& <\Theta\ 0\ |
\vartheta(t)\Theta \varphi^\dagger(-\frac{\vec x}{2},-\frac{t}{2}) \Theta^{-1}
\Theta \varphi(\frac{\vec x}{2}, \frac{t}{2})\Theta^{-1}\cr
&&\hskip 4cm + \vartheta(-t)\Theta \varphi(\frac{\vec x}{2}, \frac{t}{2})\Theta^{-1}
\Theta \varphi^\dagger(-\frac{\vec x}{2},-\frac{t}{2}) \Theta^{-1} \ |\ \Theta\ 0>
\cr
&=&<\Theta\ 0\ |
\vartheta(t)\Theta \varphi^\dagger(-\frac{\vec x}{2},-\frac{t}{2})
\varphi(\frac{\vec x}{2}, \frac{t}{2})\Theta^{-1}
+ \vartheta(-t)\Theta \varphi(\frac{\vec x}{2}, \frac{t}{2})
\varphi^\dagger(-\frac{\vec x}{2},-\frac{t}{2}) \Theta^{-1}\  |\ \Theta\
0>.\cr
&&
\end{eqnarray}
One now uses the antiunitarity (\ref{eq:antiunit2}) of the $\Theta$ operator
to get, using (\ref{eq:prop})
$$d(\vec x,t)= <0\ |\ \vartheta(t)\varphi^\dagger(\frac{\vec x}{2}, \frac{t}{2})
\varphi(-\frac{\vec x}{2},-\frac{t}{2})
+\vartheta(-t)  \varphi(-\frac{\vec x}{2},-\frac{t}{2})
\varphi^\dagger(\frac{\vec x}{2}, \frac{t}{2})\ |\ 0>= f(\vec x,t).$$

Accordingly, the sole existence of a antiunitary operator $\Theta=\Theta^{-1}$
 such that any complex scalar field transforms according
to (\ref{eq:WitLee}), and of the unicity and invariance of the vacuum by
$\Theta$ entail
\begin{equation}
d(\vec x,t) = f(\vec x,t).
\label{eq:d1}
\end{equation}

\subsubsection{$\boldsymbol{TCP}$ symmetry and the anti-diagonal elements}
\label{subsub:TCPantiWL}

By definition (see (\ref{eq:prop}))
\begin{equation}
-h(\vec x,t)
= e^{-i\delta}<0\ |\ \vartheta(t)\varphi^\dagger(\frac{\vec x}{2},\frac{t}{2})
\varphi^\dagger(-\frac{\vec x}{2}, -\frac{t}{2})
+ \vartheta(-t)\varphi^\dagger(-\frac{\vec x}{2}, -\frac{t}{2})
\varphi^\dagger(\frac{\vec x}{2},\frac{t}{2}) \ |\ 0>.
\label{eq:h0}
\end{equation}
The vacuum is supposed to be unique and invariant by $TCP$: $\Theta \ |\ 0> =
|\ 0>$; using (\ref{eq:WitLee}), one  gets by definition, along the same
lines as for the diagonal elements,
\footnote{In the equalities below, one cannot cancel the two $\Theta$'s in
$<\Theta\ 0\ |\Theta \varphi \ldots$ since it is equal to
$<0\ |\ \Theta^\dagger\Theta \varphi \ldots$, and $\Theta$ is not a unitary
operator, $\Theta^\dagger\Theta \not=1$.}
\begin{eqnarray}
-h(\vec x,t) &=& e^{-i\delta}<\Theta\ 0\ |
\vartheta(t)\Theta \varphi(-\frac{\vec x}{2},-\frac{t}{2}) \Theta^{-1}
\Theta \varphi(\frac{\vec x}{2}, \frac{t}{2})\Theta^{-1}\cr
&&\hskip 4cm + \vartheta(-t)\Theta \varphi(\frac{\vec x}{2}, \frac{t}{2})\Theta^{-1}
\Theta \varphi(-\frac{\vec x}{2},-\frac{t}{2}) \Theta^{-1} \ |\ \Theta\
0>\cr
&=&e^{-i\delta}<\Theta\ 0\ |
\vartheta(t)\Theta \varphi(-\frac{\vec x}{2},-\frac{t}{2})
\varphi(\frac{\vec x}{2}, \frac{t}{2})\Theta^{-1}
+ \vartheta(-t)\Theta \varphi(\frac{\vec x}{2}, \frac{t}{2})
\varphi(-\frac{\vec x}{2},-\frac{t}{2}) \Theta^{-1} \ |\ \Theta\ 0>.\cr
&&
\end{eqnarray}
One now uses the antiunitarity of the $\Theta$ operator (\ref{eq:antiunit2})
to get
\begin{equation}
-h(\vec x,t)= e^{-i\delta}<0\ |\ \vartheta(t)\varphi^\dagger(\frac{\vec x}{2}, \frac{t}{2})
\varphi^\dagger(-\frac{\vec x}{2},-\frac{t}{2})
 + \vartheta(-t)\varphi^\dagger(-\frac{\vec x}{2},-\frac{t}{2})
\varphi^\dagger(\frac{\vec x}{2}, \frac{t}{2})\ |\ 0> =-h(\vec x,t).
\label{eq:h1}
\end{equation}
We only get a tautology: $TCP$ sets no constraint on the antidiagonal
elements of the propagator.

\subsection{THE SCHWINGER-PAULI CONVENTION FOR $\boldsymbol{TCP}$ TRANSFORMATION
\cite{AkhiezerBerestetskii}\cite{Pauli}\cite{Landau}}
\label{subsection:SchPau}

In the Schwinger-Pauli convention, transforming a product of operators by
$TCP$ goes by performing the 4-inversion $(\vec x,t) \rightarrow (-\vec
x,-t)$, {\em not} taking neither the hermitian nor the complex conjugate of
operators, but reading all expressions from right to left instead of from
left to right (this last prescription swaps in particular ``in'' and ``out''
states).

When evaluating a scalar product, with no ``sandwiched'' operator,
this convention is identical to the condition (\ref{eq:antiunit1})
of antiunitarity of $\Theta$; indeed, it yields
$<\Theta\  A\ |\ \Theta\  B> \stackrel{Schwinger-Pauli}{=}
 <B\ |\ A> \equiv <A\ |\ B>^\ast$.

However, when a string of operators is sandwiched between the two state
vectors, one gets a result different  from the Wightman convention; they
only coincide for hermitian operators:
\begin{equation}
<\Theta\ A\ |\ {\cal O}_1{\cal O}_2 \ldots{\cal O}_n\ |\ \Theta\ B>
\stackrel{Schwinger-Pauli}{=}
<B\ |\ {\cal O}_n\ldots{\cal O}_2 {\cal O}_1\ |\ A>
\label{eq:SP1}
\end{equation}
while it would give $<B\ |\ {\cal O}^\dagger_n\ldots{\cal O}^\dagger_2
{\cal O}^\dagger_1\ |\ A>$ in the Wightman convention (see
(\ref{eq:antiunit2})). Since (\ref{eq:antiunit2}) is a direct consequence
of the antiunitarity of the $\Theta$ operator, the differences
between the two conventions are  deep and one cannot even speak of a
true antiunitary $\Theta$ for Schwinger.

The case of ``sandwiched'' scalars needs an investigation.
If one  applies the rule of simply inverting the order of
all factors in the matrix element, one gets
$\Theta\big(< A\ |\ a\ |\  B>\big) = <B\ |\ a\ |\ A> \equiv a<B\ |\ A>$.

If one instead sticks to the antiunitarity of $\Theta$, the c-number $a$
should be complex conjugated (see (\ref{eq:antilin})).

Since we already saw above that $\Theta$ cannot be considered here as a
antiunitary operator, we shall not conjugate the c-numbers and show that
this leads to a consistent result.

A {\em caveat} also exists: Pauli \cite{Pauli} always works with completely
symmetrized strings of operators, which is not our case here.

\subsubsection{Constraint set by $\boldsymbol{TCP}$ symmetry on the diagonal
elements}
\label{subsub:TCPdiaPS}

Since ``in'' and ``out'' states are both the vacuum, supposed invariant by
$TCP$, the transformed by $TCP$ of $d(\vec x,t)$ (\ref{eq:d0}) is
\begin{equation}
\Theta d(\vec x,t) = f(\vec x,t).
\label{eq:d3}
\end{equation}
Indeed, $\Theta d(\vec x,t)=
<0\ |\ \vartheta(t)\varphi^\dagger(\frac{\vec x}{2},\frac{t}{2})
\varphi(-\frac{\vec x}{2}, -\frac{t}{2})
+ \vartheta(-t)\varphi(-\frac{\vec x}{2}, -\frac{t}{2})
\varphi^\dagger(\frac{\vec x}{2},\frac{t}{2}) \ |\ 0> = f(\vec x,t)$.

A $TCP$ transformation swaps the two diagonal elements; $TCP$ invariance
requires accordingly their equality.

\subsubsection{Constraint set by  $\boldsymbol{TCP}$ symmetry on
the anti-diagonal elements}
\label{subsub:TCPantiPS}

The transformed by $TCP$ of $-h(\vec x,t)$ (\ref{eq:h0}) is, when
the phase is not transformed ---see the discussion above---
\begin{equation}
\Theta h(\vec x,t) = h(\vec x,t).
\label{eq:h4}
\end{equation}
Indeed,
$\Theta (-h(\vec x,t)) = e^{-i\delta} < 0\ |\ \vartheta(t)
\varphi^\dagger(\frac{\vec x}{2},\frac{t}{2})
\varphi^\dagger(-\frac{\vec x}{2}, -\frac{t}{2})
+ \vartheta(-t)
\varphi^\dagger(-\frac{\vec x}{2},-\frac{t}{2})
\varphi^\dagger(\frac{\vec x}{2}, \frac{t}{2})\
|\  0> = -h(\vec x,t)$,
and $TCP$ does not set any constraint on the antidiagonal elements of the
propagator

If one does conjugate the phase, the relation becomes
$\Theta h(\vec x,t) = e^{2i\delta} h(\vec x,t)$, which
is not consistent with what we obtained using the Wightman's convention.

\subsubsection{Comments}
\label{subsub:comm2}

In the Schwinger-Pauli convention for $TCP$-transforming a string of
operators, $TCP$ invariance constrains the two diagonal elements to be
identical and gives no constraint on the antidiagonal elements.

This is only achieved when c-numbers are left untouched by the
transformation.

The operator $\Theta$ then does not appear as an antiunitary operator;
nevertheless we get the same constraints as with the convention of Wightman.

\subsection{FINAL REMARKS ON $\boldsymbol{TCP}$}
\label{subsection:concTCP}

\subsubsection{Constraints on the propagator}
\label{subsub:consprop}

Both conventions lead us to the same constraints on the propagator:
its diagonal elements must be identical, while no constraint
exists on the antidiagonal elements.
The propagator is not constrained to be normal: depending whether
$|g| = |h|$ or not, it can or cannot be so.
In the core of the paper, we show that normality cannot be satisfied because
it always leads to a purely imaginary $CP$ violating parameter $\epsilon$.

The Schwinger-Pauli's convention has been claimed \cite{Pauli}
 to be valid independently of the hermiticity of the Lagrangian,
which is of concern to us here.  It only
coincides with the one of Wightman when dealing with hermitian operators,
but they deeply differ for other cases:
the $TCP$ operator $\Theta$ is in particular
 not truly antiunitary, nor truly antilinear in the
Schwinger-Pauli's convention
\footnote{It is to be noted that the demonstration that an operator should
be either unitary or antiunitary rests (see \cite{Weinberg}, vol.1, appendix
A p.91) on the conservation of probabilities and on the existence of a
complete orthogonal set of state-vectors for rays of the group of
transformation under concern. In the case under study neutral kaons are
unstable and we introduced a non-hermitian ``effective'' Lagrangian; hence,
one may question the conservation of probabilities; furthermore, a
complete set of eigenstates involves at most one true propagating state since,
at each of the physical poles, a complete set of eigenstates involves the
propagating (physical) state, and a spurious one; the two
propagating states correspond to two different $p^2$ and do not form, truly
speaking, a complete orthogonal set. A detailed re-examination of the
demonstration in our case is left for a further study.}
.

It is important to notice that the non-hermitian Lagrangian that one is led
to introduce because of kaon instability can only be considered as an
effective Lagrangian. A fundamental Lagrangian should include not only
kaons but all its decay products, and should be hermitian; then
Schwinger-Pauli and Wightman conventions coincide.

We develop the comparison between the two approaches in the next
subsection.

\subsubsection{Lagrangian versus Green's functions}
\label{subsub:LaGr}

We have chosen to study the constraints set by $TCP$ on the two-point
Green function. It is a general theorem \cite{StreaterWightman}
 that one can reconstruct the $S$
matrix of a theory from the (infinite) set of its Green functions, which
goes beyond any perturbative approach based on a given Lagrangian.

We show below what our results mean in a  Lagrangian approach.
A general quadratic Lagrangian for $K^0$ and $\overline{K^0}$ writes
\begin{eqnarray}
{\cal L}(\vec x,t) &=& a \varphi_{K^0}(\vec x,t) \varphi_{\overline{K^0}}(\vec x,t)
+ d \varphi_{\overline{K^0}}(\vec x,t)\varphi_{K^0}(\vec x,t)
+ b (\varphi_{K^0}(\vec x,t))^2 + c(\varphi_{\overline{K^0}}(\vec x,t))^2\cr
&=& a \varphi_{K^0}(\vec x,t)
e^{-i\delta}\varphi^\dagger_{K^0}(\vec x,t)
+  d e^{-i\delta}\varphi^\dagger_{K^0}(\vec x,t)\varphi_{K^0}(\vec x,t)
\cr
&& + b (\varphi_{K^0}(\vec x,t))^2
+ c e^{-2i\delta} (\varphi^\dagger_{K^0}(\vec x,t))^2.
\end{eqnarray}
If one uses the Wightman convention for $TCP$ transformation
(which includes complex conjugating the c-numbers) one gets
\begin{eqnarray}
\Theta {\cal L}(\vec x,t)\Theta^{-1} &=&
a^\ast e^{i\delta}\varphi_{K^0}(-\vec x,-t)\varphi^\dagger_{K^0}
(-\vec x,-t)
+ d^\ast e^{i\delta}\varphi^\dagger_{K^0}(-\vec x,-t)\varphi_{K^0}
(-\vec x,-t) \cr
&& + b^\ast (\varphi^\dagger_{K^0}(-\vec x,-t))^2
+ c^\ast e^{2i\delta} (\varphi_{K^0}(-\vec x,-t))^2;
\end{eqnarray}
if one uses instead the Schwinger-Pauli convention (with
no conjugation of c-numbers), one gets
\begin{eqnarray}
\Theta {\cal L}(\vec x,t)\Theta^{-1} &=&
a e^{-i\delta}\varphi^\dagger_{K^0}(-\vec x,-t)\varphi_{K^0}
(-\vec x,-t)
+ d e^{-i\delta}\varphi_{K^0}(-\vec x,-t)
\varphi^\dagger_{K^0}(-\vec x,-t) \cr
&& + b (\varphi_{K^0}(-\vec x,-t))^2
+ c e^{-2i\delta} (\varphi^\dagger_{K^0}(-\vec x,-t))^2,
\end{eqnarray}
such that, supposing  that the
integration $\int d^4x$ wipes out the change of $(\vec x,t)$ into $(-\vec
x,-t)$, the conditions that $\cal L$ is invariant by $TCP$
are:\newline\null
- in the Wightman's convention: $ae^{-i\delta}= a^\ast
  e^{i\delta}, de^{-i\delta}= d^\ast e^{i\delta},
b=c^\ast e^{2i\delta}$;\newline\null
- in the Schwinger-Pauli convention: $a=d$, no condition on $b$ and $c$:
  these are the same results as the ones that we have obtained for the
propagator (which, at the lowest order, is the inverse of the quadratic
Lagrangian)
\footnote{
When the arbitrary phase $\delta$ is put to zero, these constraints
become:\newline\null
- in the Wightman's convention: $a$ and $d$ real, $b=c^\ast$;\newline\null
- in the Schwinger-Pauli convention: $a=d$ (equality of the masses of $K^0$
and $\overline{K^0}\equiv (K^0)^\dagger$)  and no condition on $b$ and $c$.}
.

For the Lagrangian to be hermitian, $a$, $b$, $c$ and $d$ must satisfy
$ae^{-i\delta}$ real, $d e^{-i\delta}$ real, $b^\ast = ce^{-2i\delta}$;
these coincide with the
$TCP$ constraints that we obtained above by using the conventions of Wightman
\footnote{Since the Lagrangian involves operators at the same point and no
T-product is involved, the $TCP$ transformation amounts, in addition to
4-inversion, to simple hermitian conjugation.}
. What matters then is the sum $(a+d)$, as far as $\varphi_{K^0}(x)$ and
$\varphi^\dagger_{K^0}(y)$ commute at the same space-time point $x=y$
\footnote{This gives back, at the Lagrangian level, the
equality between the mass of a particle and the one of the corresponding
antiparticle.}
.

We conclude that:\newline\null
- the Schwinger-Pauli convention always gives  constraints
for the propagator and for the Lagrangian which are
compatible;\newline\null
- in the case when the instability of particles forces us to
use a non-hermitian Lagrangian,
Wightman's convention for the propagator and for the Lagrangian conflict.

So, as reported in Pauli's paper \cite{Pauli},
Schwinger's convention for $TCP$ is
likely to apply to non-hermitian Lagrangians as well, and looks more
general than Wightman's.

When dealing with non-hermitian Lagrangians, it seems thus
recommended:\newline\null
- either to work with Green's functions only;  one can then keep a
  antiunitary $\Theta$ operator and Wightman's convention;\newline\null
- or to use Schwinger-Pauli convention for $TCP$ which, in particular,
brings no conflict between Lagrangian and propagator; one has then to give up
the antiunitarity of the $\Theta$ operator.

\section{$\boldsymbol{CP}$ VIOLATING PARAMETERS AND ARBITRARY PHASES}
\label{section:epsilon}

This appendix is a complement to subsection \ref{subsub:CPparam}.

First, on Fig.~2 are displayed the two ellipsoids of $\epsilon_L^{in}$ and
$\epsilon_L^{out}$ when the arbitrary phase $\Omega_1$ in (\ref{eq:nota6}) is
varied.

\vbox{
\begin{center}
\includegraphics[height=10truecm,width=6truecm,angle=-90]{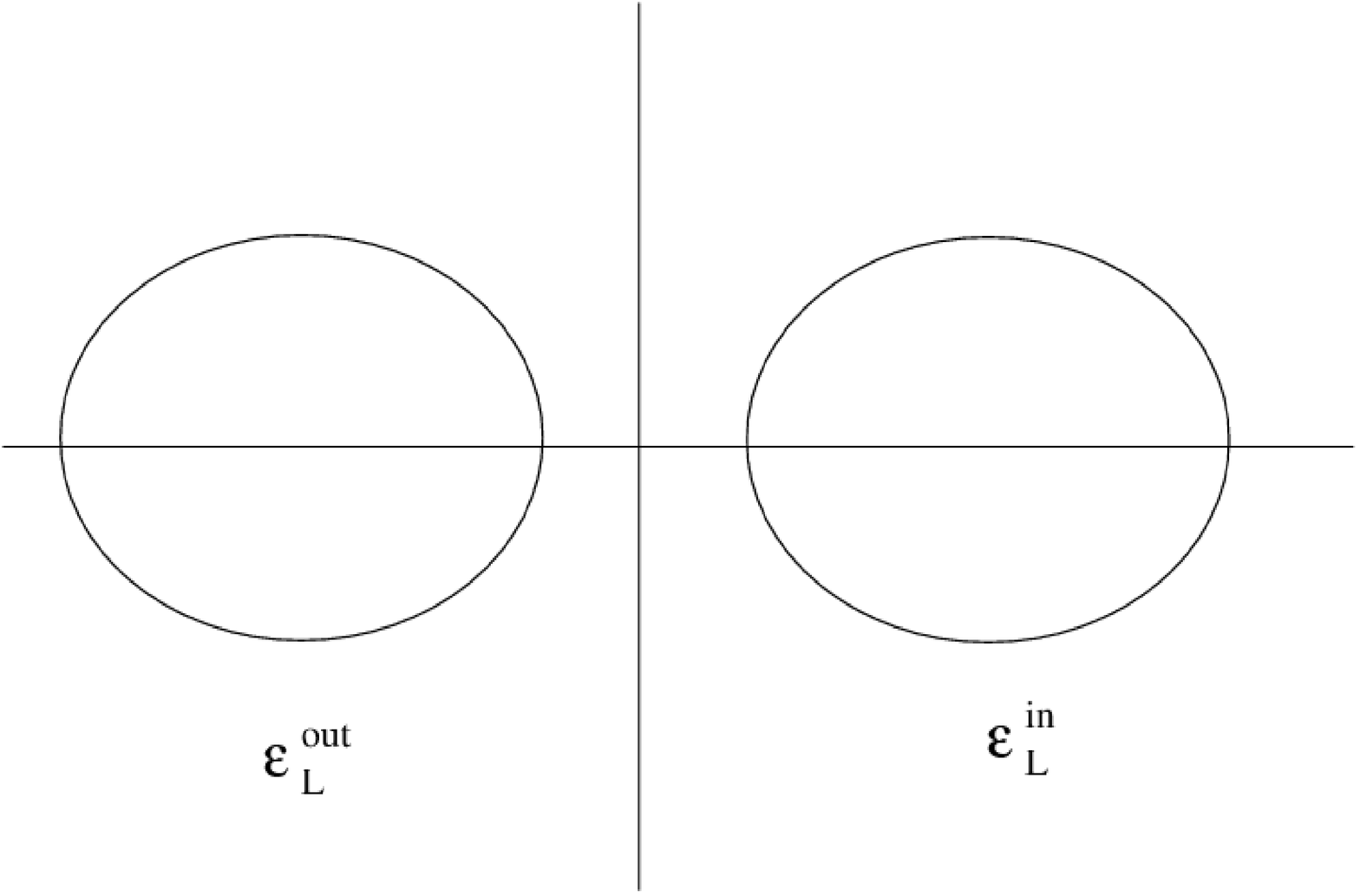}
\end{center}
\figskip
\centerline{\em Fig.~2: $\epsilon_L^{in}$ and $\epsilon_L^{out}$ span
symmetric ellipsoids when the phase $\omega$ of $K^0$  is varied.}
\figskip
}

The coordinates of their intersections with the real axis have the same
moduli as the lower and upper bounds (\ref{eq:boundmod}). The length
of their ``real'' axis of the ellipsoids is
$\left|\frac{4\sqrt{|b_1c_1|}}{|b_1|-|c_1|}\right|$
and the length of their ``imaginary'' axis is very close to $2$.
(Fig.~2 does not respect the scale:  the lower bound of $\epsilon$ is
much smaller than the length of the ``imaginary'' axis, itself much smaller
that the length of the ``real'' axis).

By the same inspection as done in subsection \ref{subsub:CPparam}, one
finds
\begin{eqnarray}
&&\epsilon_{\tilde L}^{in} = \frac
{|b(z_1)| -|c(z_1)|- 2i \sqrt{|b(z_1)c(z_1)|}\sin\Omega_1}
{|b(z_1)| +|c(z_1)|- 2 \sqrt{|b(z_1)c(z_1)|}\cos\Omega_1};\quad
\epsilon_{\tilde L}^{out} = \frac
{|c(z_1)| -|b(z_1)|- 2i \sqrt{|b(z_1)c(z_1)|}\sin\Omega_1}
{|b(z_1)| +|c(z_1)|- 2 \sqrt{|b(z_1)c(z_1)|}\cos\Omega_1};\cr
&& \cr
&&\hskip 3cm|\epsilon_{\tilde L}^{in}|^2 = |\epsilon_{\tilde L}^{out}|^2=
 \frac{|b_1| + |c_1| +2\sqrt{|b_1c_1|} \cos\Omega_1}
{|b_1| + |c_1| -2\sqrt{|b_1c_1|} \cos\Omega_1}.
\label{eq:epstilde}
\end{eqnarray}
Comparing (\ref{eq:epstilde}) with (\ref{eq:epsilons3}), one sees that
the $CP$ violating parameters for the ``spurious'' states can be deduced
from the ones of the propagating states by the change

\begin{equation}
\Omega_1 \rightarrow \Omega_1 \pm\pi,
\label{eq:reph1}
\end{equation}
that is, by 
\begin{equation}
\omega \rightarrow \omega \pm \frac{\pi}{2},
\label{eq:reph2}
\end{equation}
which is also true for $K_S$ and $K_{\overline S}$.

So, the ellipsoids corresponding to the ``spurious states'' are
globally the same as the ones of the propagating states.  The value
of $\Omega_1$ corresponding to the lower bound of $|\epsilon|$ in one case,
$\Omega_1 = 0 \ or \ \Omega_1 =\pi$, corresponds to the upper bound in
the other case.

The ellipsoids corresponding to $K_S$ are shifted with respect to the ones
for $K_L$ according to the transformations
$b(z_1) \rightarrow b(z_2)$ and $c(z_1) \rightarrow c(z_2)$,
shift expected to be very small; they have the same symmetry properties.

Since by (\ref{eq:reph2}), $\epsilon_L^{in}$ and $\epsilon_L^{out}$ are
turned into $\epsilon_{\tilde L}^{in}$ and $\epsilon_{\tilde L}^{out}$,
the propagating $K_L$ states  are formally transformed by this rephasing
 into the ``spurious'' states $K_{\tilde L}$, and {\em vice versa}.
The same remark applies to $K_S$ and $K_{\tilde S}$.
This is an additional argument for the importance of
both types of states, and that discarding {\em a priori} the
spurious states is unjustified.


\newpage\null
\listoffigures
\bigskip
\begin{em}
Fig.~1: Feynman diagrams for the transition $K^0 \rightarrow
K^0$;\newline\null
Fig.~2: $\epsilon_L^{in}$ and $\epsilon_L^{out}$ span
symmetric ellipsoids when the phase $\omega$ of $K^0$  is varied.
\end{em}


\newpage
%
%
%
\begin{em}

\end{em}

%
\end{document}